\documentclass{article}

\usepackage[preprint, nonatbib]{neurips_2024}
\usepackage[utf8]{inputenc} 
\usepackage[T1]{fontenc}    
\usepackage{hyperref}       
\usepackage{url}            
\usepackage{booktabs}       
\usepackage{amsfonts}       
\usepackage{nicefrac}       
\usepackage{microtype}      
\usepackage{xcolor}         
\usepackage{multirow}
\usepackage{makecell}
\usepackage{algorithm}
\usepackage{algpseudocode}
\usepackage{graphicx}
\usepackage{float}
\usepackage{stfloats}
\usepackage{amsmath}
\usepackage{amsthm}
\usepackage{amssymb}
\usepackage{amsfonts} 
\usepackage{mathtools}

\usepackage{caption}
\usepackage{subcaption}

\newtheorem{theorem}{Theorem}[section]

\algrenewcommand\algorithmicrequire{\textbf{Input:}}
\algrenewcommand\algorithmicensure{\textbf{Output:}}

\title{Exploring Loss Landscapes through the Lens of Spin Glass Theory}

\author{
    \textbf{Hao Liao $^1$} \quad ~\textbf{Wei Zhang $^1$} \quad ~\textbf{Zhanyi Huang $^1$} \quad ~\textbf{Zexiao Long $^1$} \\ 
    \textbf{Mingyang Zhou $^1$} \quad ~\textbf{Xiaoqun Wu $^1$} \quad ~\textbf{Rui Mao $^1$} \quad ~\textbf{Chi Ho Yeung $^2$}\thanks{Chi Ho Yeung is the corresponding author, email: chyeung@eduhk.hk}\\
    \\
	$^{1}$ College of Computer Science and Software Engineering, Shenzhen University \\ $^{2}$  Department of Science and Environmental Studies, The Education University of Hong Kong \\
    {\tt\small\{haoliao, zmy, mao\}@szu.edu.cn, }\\
    {\tt\small\{2210275010, huangzhanyi2020, 2019152104, mao\}@email.szu.edu.cn, }\\
    {\tt\small\{xqwu\}@whu.edu.cn}  \\
    {\tt\small\{chyeung\}@eduhk.hk}}

\begin{document}

\maketitle

\begin{abstract}
In the past decade, significant strides in deep learning have led to numerous groundbreaking applications. Despite these advancements, the understanding of the high generalizability of deep learning, especially in such an over-parametrized space, remains limited. For instance, in deep neural networks (DNNs), their internal representations, decision-making mechanism, absence of overfitting in an over-parametrized space, superior generalizability, etc., remain less understood. Successful applications are often considered as empirical rather than scientific achievement. This paper delves into the loss landscape of DNNs through the lens of spin glass in statistical physics, a system characterized by a complex energy landscape with numerous metastable states, as a novel perspective in understanding how DNNs work. We investigated the loss landscape of single hidden layer neural networks activated by Rectified Linear Unit (ReLU) function, and introduced several protocols to examine the analogy between DNNs and spin glass. Specifically, we used (1) random walk in the parameter space of DNNs to unravel the structures in their loss landscape; (2) a permutation-interpolation protocol to study the connection between copies of identical regions in the loss landscape due to the permutation symmetry in the hidden layers; (3) hierarchical clustering to reveal the hierarchy among trained solutions of DNNs, reminiscent of the so-called Replica Symmetry Breaking (RSB) phenomenon (i.e. the Parisi solution) in spin glass; (4) finally, we examine the relationship between the ruggedness of DNN's loss landscape and its generalizability, showing an improvement of flattened minima.

\end{abstract}

\section{Introduction}

The rapid development of deep learning in the past decade has inspired a lot of extraordinary applications and achieved remarkable success in various fields, from machine vision \cite{DBLP:conf/iccv/LiuL00W0LG21}, natural language processing \cite{10.1145/3649506}, to psychology \cite{DBLP:conf/acl/ZhuP0ZH20} and education \cite{DBLP:conf/aaai/AgrawalPDLC24}. Yet, comprehension of the underlying mechanism of its exceptional performance remains limited. For instance, the internal representations of deep neural networks (DNNs), the mechanisms by which they achieve effective decision-making, the absence of overfitting in over-parameterized DNNs, and their superior generalizability, are not fully understood. As a result, the success of DNNs remains an empirical rather than a scientific achievement. This presents both exciting challenges and opportunities to develop a comprehensive understanding of DNNs. 
Moreover, since deep learning is now commonly employed in important applications such as medical image analysis \cite{DBLP:conf/iccv/ButoiOMSGD23} and autonomous vehicles \cite{10614862}, this incomplete understanding can have potentially severe consequences. 
Therefore, it is imperative to gain a deeper understanding of the working principles of deep neural networks (DNNs) in order to optimize their performance and mitigate the associated risks.

Indeed, attempts from various areas have been made to understand DNNs \cite{ DBLP:journals/corr/GoodfellowSS14, DBLP:journals/cacm/ZhangBHRV21, DBLP:conf/nips/ZhouF0XHE20}. 
One line of research employed statistical physics has been an effective tool to investigate shallow artificial neural networks (ANNs) \cite{DBLP:conf/colt/AbbeAM22, DBLP:conf/nips/AdlamP20, DBLP:conf/colt/0002SKL23}.
Unlike the approaches in computer science, statistical physics offers a bottom-up perspective to explain macroscopic behavior based on microscopic interactions  \cite{carleo2019machine, DBLP:journals/entropy/GostiFLR19}. However, due to the complexity inherent in DNNs, existing research employing statistical physics tools mainly focused on shallow networks \cite{NEURIPS2022_5b2db6df, DBLP:conf/icml/KarakidaTHO23, DBLP:conf/icml/Arora0P22}, networks with random weights \cite{NEURIPS2021_e53a0a29, li2018exploring} or networks with random data \cite{NEURIPS2022_af5509c7}. 
Although researchers have endeavored to leverage statistical physics for understanding DNNs, less realistic settings or assumptions were often made to make theoretical analyses feasible. Consequently, many mysteries surrounding DNNs remain unsolved.

To this end, we analyze the loss landscape of one-hidden-layer neural networks with ReLU (Rectified Linear Unit) activation functions to obtain a fundamental understanding of their superior performance and apply this understanding to practical applications. Specifically, we apply statistical physics approaches to achieve a more in-depth understanding on the loss landscape of DNNs by drawing analogy with spin glass. We attempt to establish a framework for analysis of DNNs, which can be extended to the explore DNNs with different architectures and settings. Implications are then drawn between the loss landscape and the training dynamics as well as the generalizability of DNNs, which further provide insights to their applications.

The contributions of this paper are as follows.

\begin{itemize}
    \item  We studied \textbf{random trajectories} over the loss landscape of single-hidden-layer neural networks with ReLU activation using a random walk protocol in the parameter space, examining these trajectories to identify loss or entropy barriers.
    \item  The \textbf{permutation symmetry} of nodes in DNN hidden layers can create identical regions in the loss landscape, leading to multiple extrema. We investigated this by introducing a permutation-interpolation protocol, which swaps nodes and their link weights in the hidden layer to generate different DNNs with identical outputs, placing them in different copies of these identical regions.
    \item We analyzed various trained DNN solutions using a \textbf{hierarchical clustering} algorithm, similar to the analysis of low-energy states in spin glass. This revealed a similarity between the hierarchical structure of trained DNN solutions and the spin glass model, particularly in comparison to the Replica Symmetry Breaking (RSB) phase, which exhibits complex and interesting behavior.
    \item We revealed the dependence of DNNs' generalizability on the \textbf{flatness of extrema} in the loss landscape and the batch size used in gradient descent during training.
\end{itemize}

\section{Related Works}

\subsection{The Loss Landscape of DNNs}

Deep neural networks (DNNs) are extensively utilized across various domains \cite{DBLP:conf/iccv/LiuL00W0LG21, DBLP:conf/icml/MandalTR23}, prompting significant efforts to understand their underlying mechanisms \cite{goodfellow2016deep}. 
Despite being overparameterized, DNNs tend to generalize well with unknown data and rarely suffer from overfitting or poor local minima. 
This phenomenon is largely attributed to the complex loss landscape in the weight parameter space. Thus the loss landscape plays a critical role in understanding DNNs. \cite{DBLP:journals/frai/SahsPDCTLAP22, DBLP:conf/nips/Li0TSG18}
The flatness of minima is often closely related to the generalization of DNNs. Some researchers posit that flat minima contribute positively to generalization \cite{DBLP:conf/nips/ZhangZLX21, DBLP:conf/iclr/KeskarMNST17}
However, others contend that measures of flatness lack robustness due to parameter symmetries, and that re-parameterizing weights can enhance generalization even to non-flat minima \cite{DBLP:conf/icml/DinhPBB17, DBLP:conf/nips/BoursierPF22}. 
Moreover, studies have shown that under certain limited assumptions, the loss landscape contains many degenerate global minima, which the training algorithm is likely to locate \cite{DBLP:conf/nips/Kawaguchi16, DBLP:conf/nips/0001E023}. Despite these efforts to understand the generalization of DNNs through the loss landscape, conclusive results have yet to be achieved.

Analyzing the loss landscape helps understand DNNs' training dynamics.  Although training DNNs constitutes an NP-hard non-convex optimization problem \cite{DBLP:conf/nips/BlumR88, DBLP:conf/nips/WuX020}, some studies have shown that the probability of a monotonically decreasing path from the initial weight configuration to the global optimum is high \cite{DBLP:conf/icml/SafranS16}. Others argue that proper initialization of Gradient Descent (GD) leads to implicit regularization \cite{zhang2017musings}, and that Stochastic Gradient Descent (SGD) may identify one of many degenerate minima \cite{DBLP:journals/corr/PoggioL17}. SGD based on local entropy has been introduced to leverage the conjecture that good minima are characterized by flatness \cite{DBLP:conf/nips/ZhouF0XHE20}. 
Additionally, analyzing loss landscapes has been used to enhance the robustness of training algorithms against adversarial examples, thereby reducing the risk of misclassification \cite{DBLP:conf/nips/WuX020, DBLP:journals/corr/GoodfellowSS14}. Thus, a more fundamental and comprehensive understanding of the loss landscape can improve model performance and mitigate the risks associated with deep learning applications.

\subsection{Spin Glass Theory}

The statistical physics community has a long history of researching artificial neural networks (ANNs) \cite{DBLP:conf/icml/SutskeverMDH13, DBLP:conf/nips/SorscherGSGM22}. ANNs are often analyzed using methodologies developed in the studies of spin glass \cite{mezard1987spin, nishimori2001statistical}, as the training of ANNs is also an optimization problem subject to a set of fixed input and outputs, similar to identifying the group state in spin glass with quenched coupling disorders \cite{hertz2018introduction}. Early advancements in understanding ANNs through statistical physics include studies on associative memory in Hopfield models \cite{hopfield1982neural, amit1985storing}, perceptron storage capacity and generalization \cite{gardner1987maximum, gardner1988space}, and learning dynamics in perceptrons and two-layer neural networks, such as committee machines \cite{saad1995line, DBLP:conf/nips/FreundSST92}. 

The statistical physics community has a long history of researching artificial neural networks (ANNs) using methods from spin glass studies. \cite{DBLP:conf/icml/SutskeverMDH13, DBLP:conf/nips/SorscherGSGM22} Training ANNs is an optimization problem similar to identifying the group state in spin glass with quenched coupling disorders. \cite{mezard1987spin, nishimori2001statistical} Early advancements in understanding ANNs through statistical physics include studies on associative memory in Hopfield models, \cite{hopfield1982neural, amit1985storing} perceptron storage capacity and generalization, \cite{gardner1987maximum, gardner1988space} and learning dynamics in perceptrons and two-layer neural networks, such as committee machines. \cite{saad1995line, DBLP:conf/nips/FreundSST92}

Unlike conventional studies, which focuses on performance and solving specific instances, the use of physics tools aims to reveal the the general underlying mechanisms of ANNs. Recently, the statistical physics community has sparkled a new wave of focus on DNN analysis, although assumptions such as random weights or random data are often employed to facilitate theoretical analyses \cite{huang2015advanced, li2018exploring}. These analytical physical insights often inspire practical applications. For instance, Onsager reactions have been incorporated to improve DNN training \cite{DBLP:journals/pnas/BaldassiBCILSZ16}, and robust ensembles describing well-configured density regions have inspired new entropy-based training algorithms \cite{DBLP:conf/iclr/ChaudhariCSLBBC17, DBLP:conf/nips/GabrieMLBMKZ18}. Clearly, statistical physics tools continue to contribute significantly to the fundamental understanding of DNNs and their applications.

\section{Methods}

\subsection{Random Walk and Interpolation on the Loss Landscape}

\begin{algorithm}
\caption{MCMC random walk algorithm}
\label{alg:MCMC}
\begin{algorithmic}[1]
\Require Random walk steps $T$, Gaussian distribution $\mathcal{N}(0, \sigma^2)$
\Ensure New weight parameter configuration $\vec{\textbf{w}}'_{trained}$, loss and accuracy
\State Generate trained parameter configurations $\vec{\textbf{w}}_{trained}$
\State Start with $\vec{\textbf{w}}(t=0)=\vec{\textbf{w}}_{trained}$
\For{$i = 1$ to $T$}
\State Draw randomly a vector $\Delta\vec{\textbf{w}}(t)$ from distribution $\mathcal{N}(\Delta\vec{\textbf{w}})$
\State $\vec{\textbf{w}}(t)=\vec{\textbf{w}}(t-1)+\Delta\textbf{w}(t)$
\State Compute loss and accuracy
\EndFor
\State \Return $\vec{\textbf{w}}^{\prime}_{trained}$, loss, and accuracy
\end{algorithmic}
\end{algorithm}

The loss landscape of DNNs is highly complex and non-convex, often containing multiple local extrema or saddle points. These features lead to different trained solutions depending on initialization conditions, which can adversely affect the model's generalizability.
\textbf{Random walks} can be used to prob the parameter space from the different weight configurations. 
Here, we use $\vec{\textbf{w}}$ to denote the vector of parameters in the DNN. The random walk starts from different initial parameter configurations $\vec{\textbf{w}}_{trained}$ obtained by training the DNN on specific dataset. For each time step $t$, we draw randomly a vector $\Delta\vec{\textbf{w}}(t)$ from a distribution $\mathcal{N}(\Delta\vec{\textbf{w}})$. The parameter configuration at time $t+1$ is defined in Eq.~\ref{eq_rw}. 

\begin{equation}
\label{eq_rw}
    \vec{\textbf{w}}_{t+1} = \vec{\textbf{w}}_{t} + \Delta\vec{\textbf{w}}(t)
\end{equation}

This process follows a \emph{Markov Chain Monte Carlo (MCMC)} procedure that are detailed in Algorithm \ref{alg:MCMC}. 
As the parameter configuration randomly move away from the initial trained solution $\vec{\textbf{w}}_0$, evaluation of the loss function for each $\vec{\textbf{w}}_t$ changes accordingly, revealing the landscape along a random path in the parameter space.

Besides random walks, another way to probe the loss landscape is by exploring the \textbf{interpolation} between weight configurations, as the method introduced by Goodfellow \cite{DBLP:journals/corr/GoodfellowV14}. It relies on the linear interpolation between two trained parameter configurations $\vec{\textbf{w}}_1$ and $\vec{\textbf{w}}_2$, i.e. two points in the loss landscape, such that a point $\vec{\textbf{w}}(\alpha)$ lying on the straight line between them is given by

\begin{equation}
    \vec{\textbf{w}}(\alpha) = \alpha\vec{\textbf{w}}_1+(1-\alpha)\vec{\textbf{w}}_2,~\forall\alpha\in[0,1]
\end{equation}

where $\alpha$ controls the movement on this straight line. To probe the loss landscape, one may employ different settings for the two trained solutions $\vec{\textbf{w}}_1$ and $\vec{\textbf{w}}_2$. In the subsequent analysis, we will introduce a protocol to create $\vec{\textbf{w}}_2$ based on $\vec{\textbf{w}}_1$ by swapping nodes in the hidden layers of the DNNs.

\subsection{Replica Symmetry Breaking (RSB)}

\begin{figure}[!t]
    \centering
    \includegraphics[width=\linewidth]{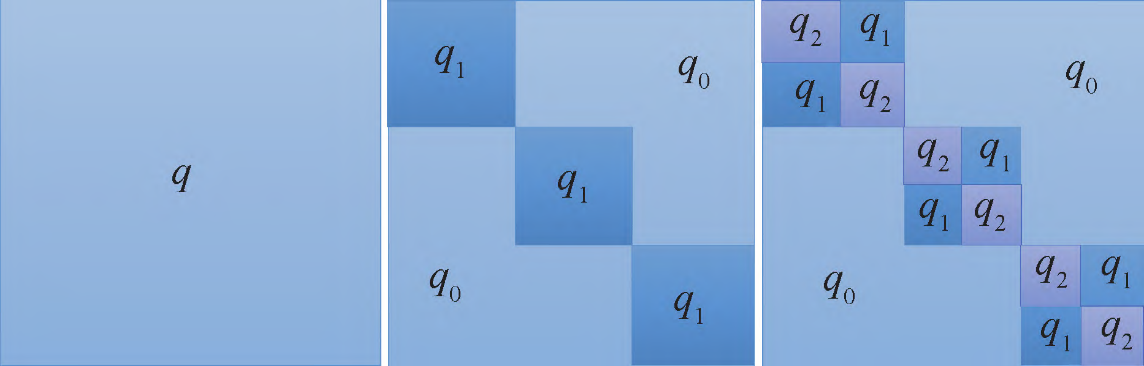}
    \caption{Different degrees of Replica Symmetry Breaking (RSB). The middle one is first-order RSB and the right one is the second-order RSB.}
    \label{fig:rsb}
\end{figure}

As a disordered system exhibiting interesting and long-lasting metastability, spin glass has its important and unique research value in statistical physics. 
Many analytical techniques were established to specifically analyze spin glass. These techniques provide new tools and perspectives for understanding information processing problems like neural networks. 
The Sherrington-Kirkpatrick (SK) model is one representative model of spin glass, which has a fully-connected structure.  
The Hamiltonian of the SK model of spin glass, analogous to the loss function in DNNs, is given by

\begin{equation}
H \propto \sum_{(ij)} J_{ij} s_i s_j,
\end{equation}

where $i$, $j$ represent the spins, $J_{i,j}$ drawn from the Gaussian distribution is the coupling strength, and $s_i=\pm 1$ represents the state.

The concept of Replica Symmetry Breaking (RSB) introduced by Parisi \cite{parisi1979infinite} offers a new perspective on the analytical solution to the SK model, which is effectively a way to characterize the structure of the energy landscape with numerous local minima. Figure \ref{fig:rsb} illustrates the case of replica symmetry, and two different degrees of RSB, namely, first-order RSB (1RSB) and second-order RSB (2RSB), which translate to an increasing degree of ruggedness in the energy landscape. 
We will analyze the loss landscape of DNNs in analogous to the structure of energy landscape in spin glass in subsequent discussion.

\begin{algorithm}
\caption{Hierarchical clustering algorithm}\label{alg:HCA}
\begin{algorithmic}[1]
\Require Solution sample set $D=\{x_1,x_2,\cdots,x_n\}$, $n$ is the number of samples, cluster distance calculation function $d$, number of clusters $k$
\Ensure Link matrix $M$, leaf node ID list, $k$ clusters $C_1,C_2,\cdots,C_k$
\For{$i=1$ to $n$}
\State $C_i=\{x_i\}$
\EndFor
\For{$i=1$ to $n$}
\For{$j=1$ to $n$}
\State $M(i,j)=d(C_i,C_j)$
\State $M(i,j)=M(j,i)$
\EndFor
\EndFor
\State $q=n$
\While{$q > k$}
\State Find the closest clusters $C_{i^*}$ and $C_{j^*}$
\State $C_{i^*}=C_{i^*}\cup C_{j^*}$
\State $q=q-1$
\For{$j=j^*+1$ to $q$}
\State remark $C_j=C_{j+1}$
\EndFor
\State delete $j^*$-th row and column in $M$
\For{$j=1$ to $q$}
\State $M(i^*, j)=d(C_{i^*}, C_j)$
\State $M(j, i^*)=M(i^*, j)$
\EndFor
\EndWhile
\end{algorithmic}
\end{algorithm}

\subsection{Permutation Symmetry and the Distance among Parameter Configurations}

Many feedforward neural networks possess the property that interchanging two units in the hidden layers and the associated link weights does not alter the network's output from specific input. Such networks are termed \emph{permutation-invariant neural network}, and this property implies a symmetrical parameter space, i.e. copies in identical regions in the loss landscape, facilitating the search for suitable weights for specific applications. 

The set of all weights in a permutation network can be denoted as the weight vector $\vec{\textbf{w}}$, with $\vec{\textbf{w}}\in \mathbb{R}^q$, where $q$ is the total number of real-valued adjustable parameters. The set of all permutations of $\vec{\textbf{w}}$ forms the non-Abelian group $S_q$ (the $q$-th symmetry group), referred to as $S$. This group contains $q!$ elements. For a permutation neural network, $T$ is defined as the set of all non-singular linear weight transformations. And for any permutation neural network $T$ is a subset of the general linear group $GL(q,R)$, which comprises all reversible linear transformations from $\mathbb{R}^q$ to $\mathbb{R}^q$. Based on these definitions, Theorem \ref{thm:rpl} can be established:

\begin{theorem}
\label{thm:rpl}
If $T$ is a proper subgroup of $GL(q,R)$, then the number of elements in $T$ satisfies $\#T \geq\prod_{i=1}^K M_i!$ , where $M_i$ represents the number of neural units in the $i$-th hidden layer, and $K$ represents the number of hidden layers.
\end{theorem}

The proof of Theorem \ref{thm:rpl} can be find in our supplementary material.

Other than the permutated solutions, multiple close-to-optimal solutions should naturally exist similar to the energy landscape of spin glass. 
To better understand the relationship among these close-to-optimal solutions, we utilize the Hierarchical Clustering Algorithm to group and identify the hierarchy among the solutions. This algorithm uses a "bottom-up" clustering strategy and produces a tree, called a {\bf Dendrogram Graph}.
There are two ways to compute the distance between two parameter configuration in parameter space. The first one is the \textbf{Euclidean distance} given by 

\begin{equation}
    \mathrm{Dist}(\vec{\textbf{w}}^1,\vec{\textbf{w}}^2):=\sqrt{\sum_{i=1}^n(w_i^1-w_i^2)^2}
\label{eq_distance}
\end{equation}

The second way is to compute the \textbf{cluster distance} between the clusters that the two points belong to, namely $d(u,v)$ given by

\begin{align}
    &d(u,v)= \nonumber \\ &\sqrt{\frac{\vert v\vert+\vert s\vert}{T}d^2(v,s)+\frac{\vert v\vert+\vert t\vert}{T}d^2(v,t)-\frac{\vert v\vert}{T}d^2(s,t)}
\end{align}
    
where $u$ is a new cluster formed by merging clusters $s$ and $t$, and $v$ is an unused cluster; $\vert u\vert$ denotes the number of elements in cluster $u$ and $T=\vert v\vert+\vert s\vert+\vert t\vert$. 

The computational complexity of the Hierarchical Clustering Algorithm \ref{alg:HCA} is $O(n^3)$, where $n$ is the number of configurations. This algorithm is a deterministic clustering algorithm and is guaranteed to converge to the target cluster number $k$.

\section{Experiments}

\subsection{Experimental Setup}

\paragraph{Datasets}

\paragraph{Datasets}
We employed four datasets of image classification widely used for benchmarking DNNs, including MNIST, fashion-MNIST, CIFAR-10 and CIFAR-100. These are all widely-used image classification dataset. The detailed information of these four datasets is in our supplementary material.

\paragraph{Network Architecture}
The Deep Neural Networks (DNNs) in this study consist of three layers. Specifically, (1) the input layer dimensions for $FC_1$ and $FC_2$ are $784$ and $3,072$, respectively. (2) The hidden layers consist of $64$, $128$, $256$, $512$, and $1024$ nodes and are activated by the ReLU function. (3) The output layer is activated by the Softmax function.

\begin{figure}
    \centering
    \includegraphics[width=\linewidth]{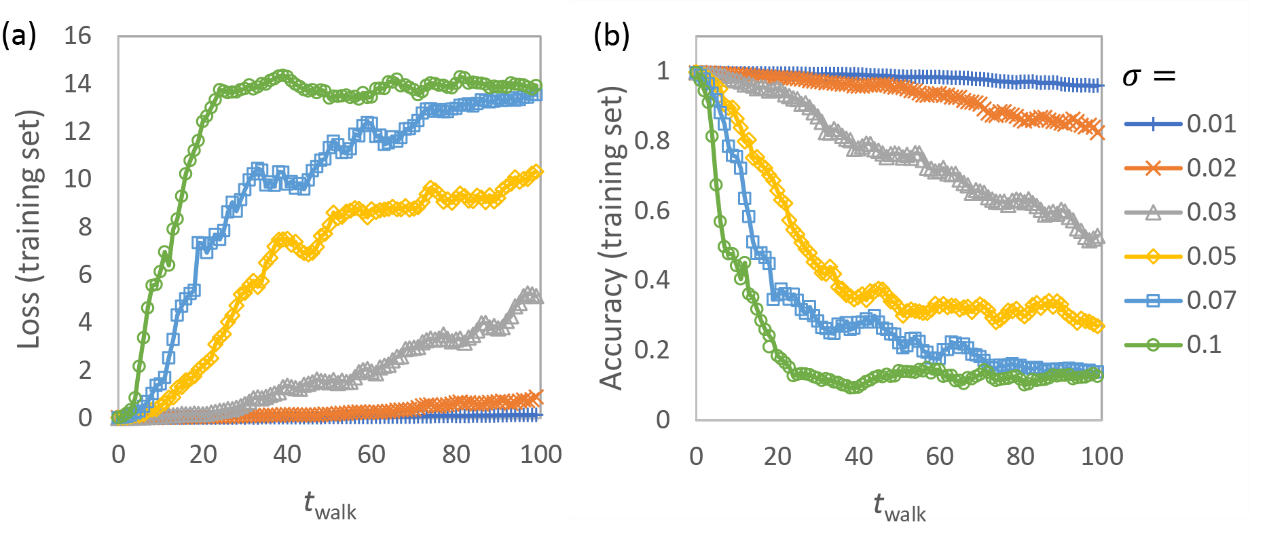}
    \caption{Examples of random walks in the parameter space from solutions $\vec{\textbf{w}}_{trained}$ of $FC_1-512$ trained on MNIST dataset by TensorFlow Adam optimizer. The parameter changes $\Delta\vec{\textbf{w}}$ are drawn from a distribution $P(\Delta\vec{\textbf{w}})=\mathcal{N}(0, \sigma^2)$, with $\sigma$ shown in the legend; the original trained parameter configuration has a magnitude  $\vert\vec{\textbf{w}}_{trained}\vert\approx 0.1$.}
    \label{fig:random_walk}
\end{figure}

\begin{figure*}[t]
	\small
	\centering
        \subfloat[ReLU, $M=5$ ]{\includegraphics[width=0.33\linewidth]{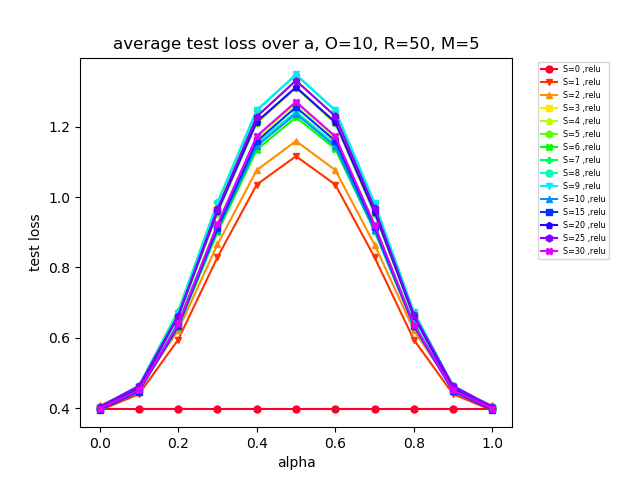}}
        \subfloat[ReLU, $M=10$]{\includegraphics[width=0.33\linewidth]{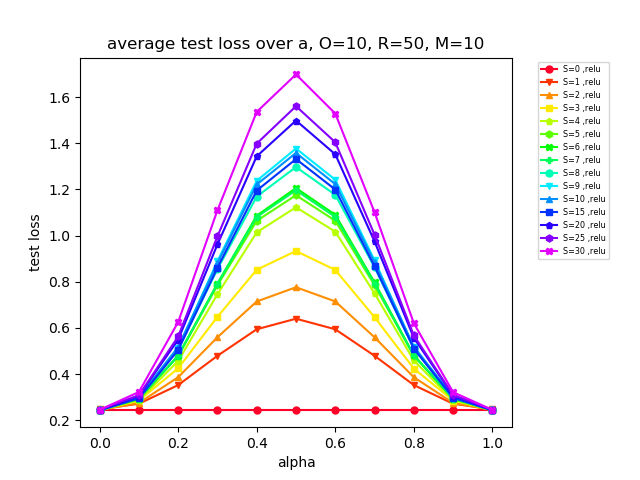}}
        \subfloat[ReLU, $M=20$]{\includegraphics[width=0.33\linewidth]{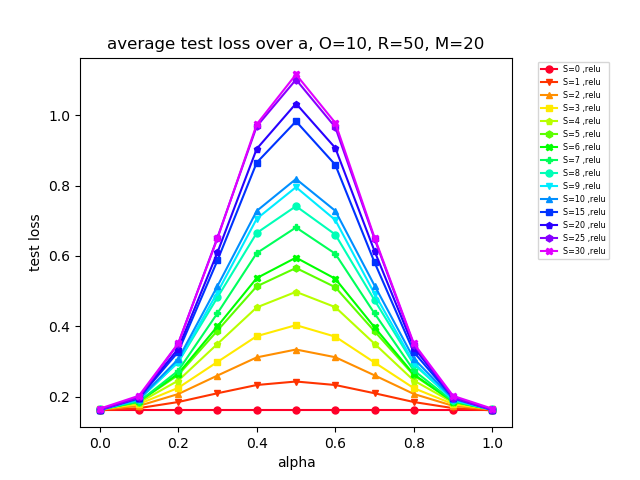}}
        \quad
	\subfloat[Softmax, $M=5$]{\includegraphics[width=0.33\linewidth]{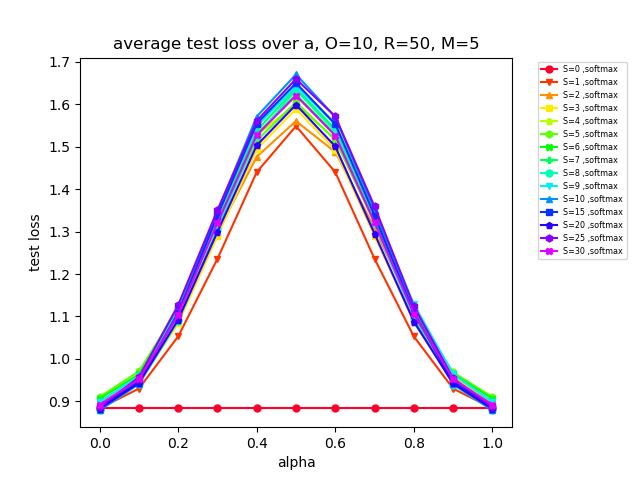}}
        \subfloat[Softmax, $M=10$]{\includegraphics[width=0.33\linewidth]{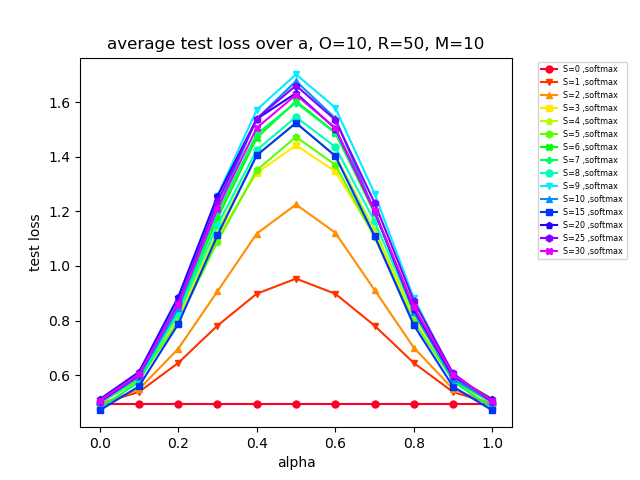}}
        \subfloat[Softmax, $M=10$]{\includegraphics[width=0.33\linewidth]{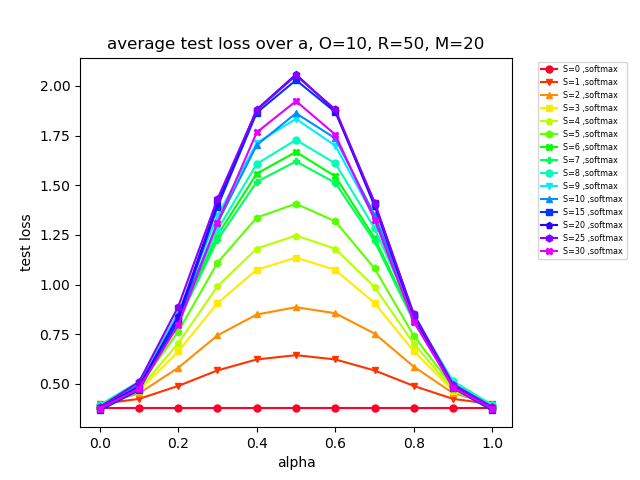}}
\caption{The test loss on the line interpolated between the original configuration and the one with swapped neuron in the hidden layer. $M$ denotes the number of nodes in the middle layer. For the output layer, the number of nodes is $10$. The activation functions are ReLU and Softmax. Results are averaged over $50$ trials. The number of swaps $S$, varies as follows:$0, 1, 2, 3, 4, 5, 6, 7, 8, 9, 10, 15, 20, 25, 30$}
\label{fig:exp_res2}
\end{figure*}

\subsection{Random Walks on the Parameter Space}

To obtain $\vec{\textbf{w}}_{trained}$, we utilized the Adam optimization algorithm to train the neural networks.
Subsequently, we employed Algorithm \ref{alg:MCMC} to generate new weight configurations. This involves conducting a new round of training and subsequently documenting the resulting training loss and accuracy. 

Figure \ref{fig:random_walk} shows the dependence of training loss and the training accuracy on $t_{walk}$, i.e. the time step of random walk. Similar trends were observed in the test set. 
As expected, the loss function increases more rapidly in cases with large $\sigma$, but the results also revealed some interesting behaviors during the
random walks, e.g. the walks traverse an initial flat region at smaller $t_{walk}$ values, potentially indicative of the minima's width, and reach a plateau of high loss when $\sigma=0.05$ and $\sigma=0.07$.These preliminary results hint at the potential to extend this approach to more complex DNNs with multiple hidden layers.

\subsection{Permutation interpolation}

According to Theorem \ref{thm:rpl}, swapping two neurons and their associated links in the hidden layer multiple times produces a new parameter configuration $\vec{\textbf{w}}$, but with the same loss and accuracy. 
We first swap neurons in the hidden layers by $S$ times, which we call the \emph{permutation} stage. Subsequently, the loss landscape was examined through interpolating between the original configurations and the transformed configurations, which we call the \emph{interpolation} stage.

Figure \ref{fig:exp_res2} shows the dependence of test loss on DNN architecture and the number of neuron swap. With the same number of swap, an increase in the depth of the hidden layer results in a smaller loss barrier at the intermediate peak. Conversely, with the same number of hidden layer nodes, more swaps results in a higher loss barrier. These results indicate that more swaps reduce the similarity between the original and the new parameter configurations, enlarging the distance between them in the loss landscape and increasing the loss barrier. In contrast, increasing the number of hidden layers tends to smooth the loss landscape, resulting in a flatter surface and a lower loss barrier.

\subsection{Loss Landscape Optimization}

\begin{figure*}[t]
	\centering
        \subfloat[Adam-256, $\mu=75.59$, $\sigma=1.27$]{\includegraphics[width=0.3\linewidth]{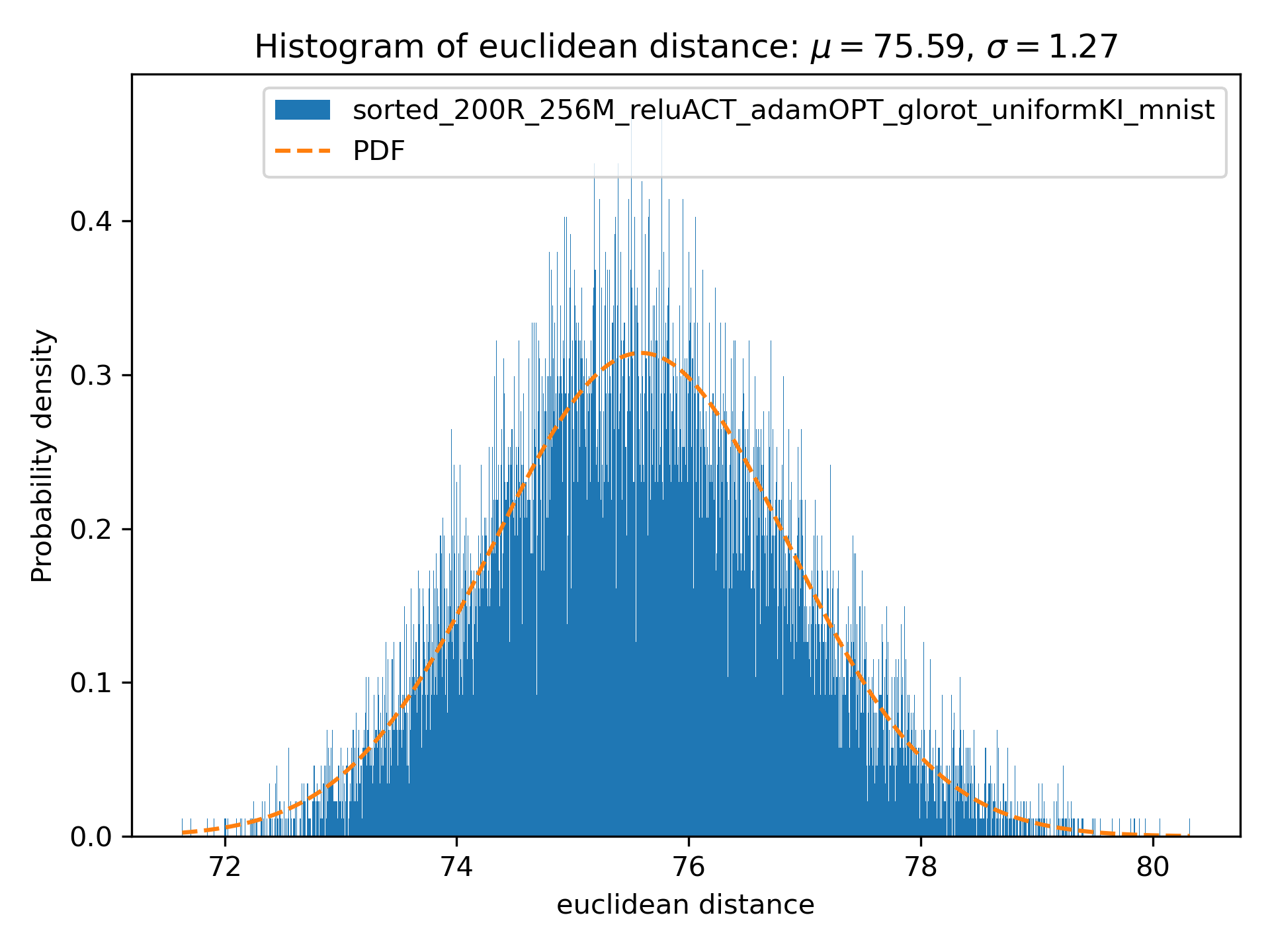}}
	\subfloat[Adam-512, $\mu=80.08$, $\sigma=2.26$]{\includegraphics[width=0.3\linewidth]{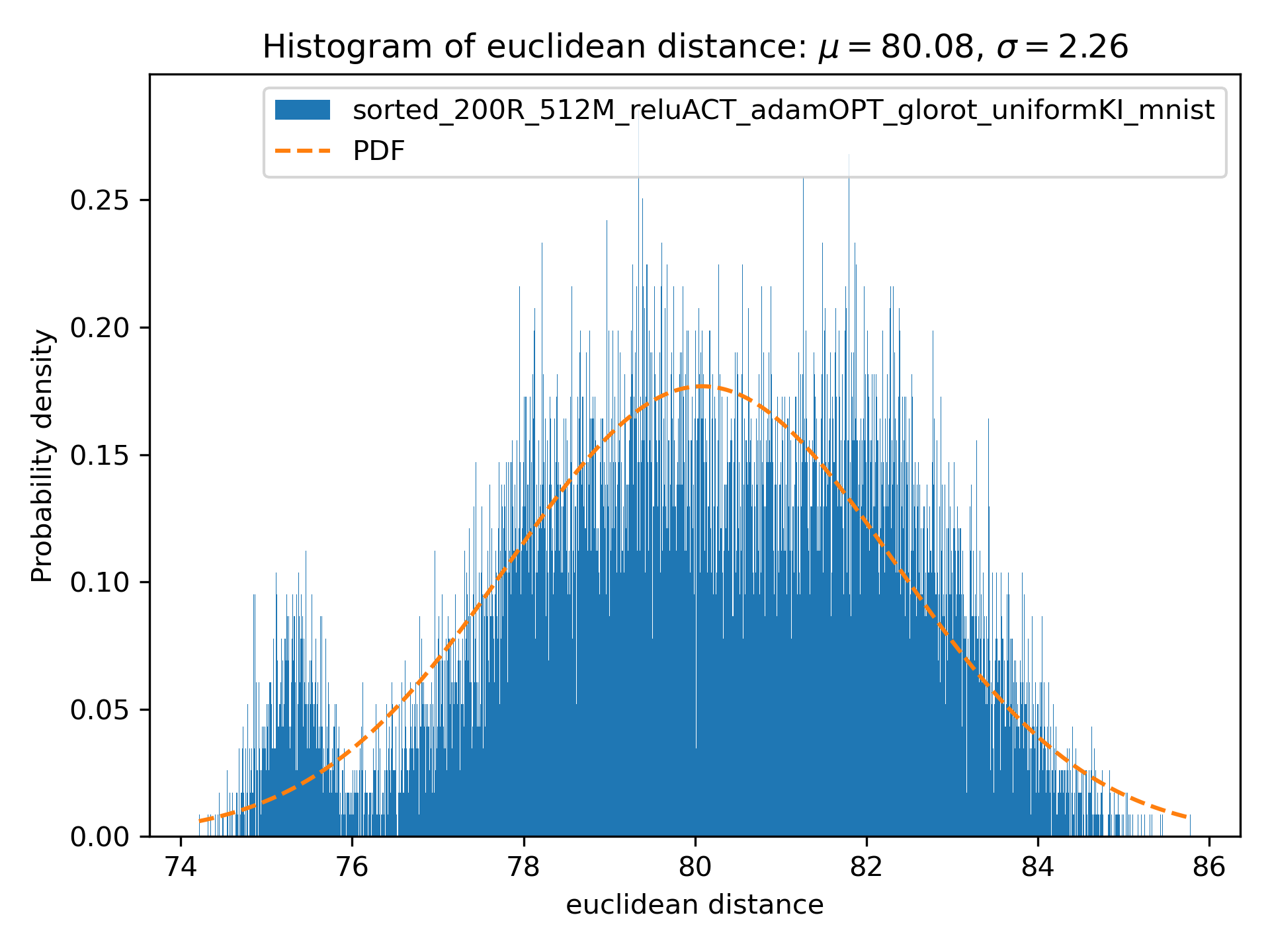}}
        \subfloat[Adam-1024, $\mu=86.26$, $\sigma=2.26$]{\includegraphics[width=0.3\linewidth]{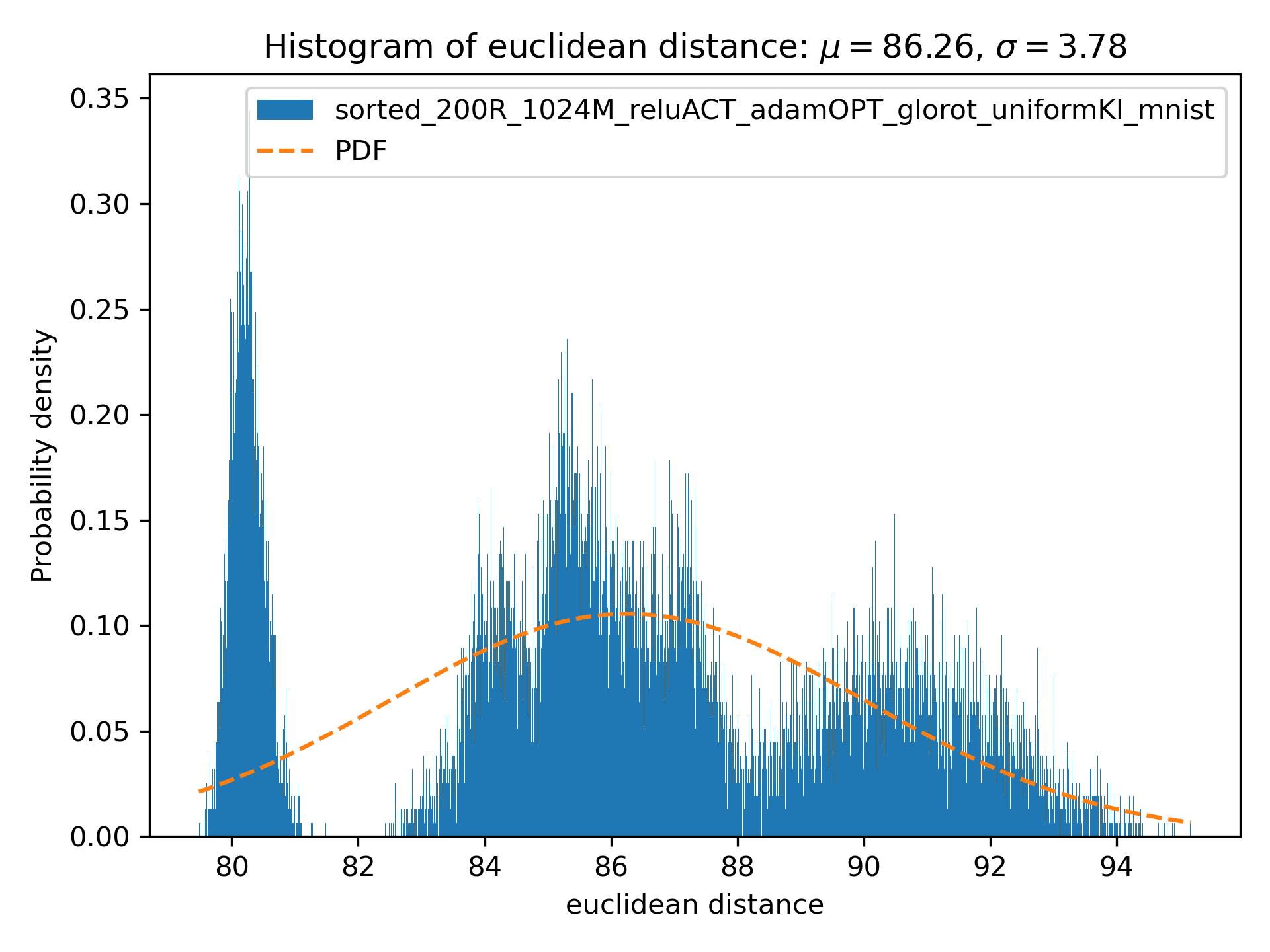}}
        \quad
        \subfloat[SGD-256, $\mu=40.38$, $\sigma=0.17$]{\includegraphics[width=0.3\linewidth]{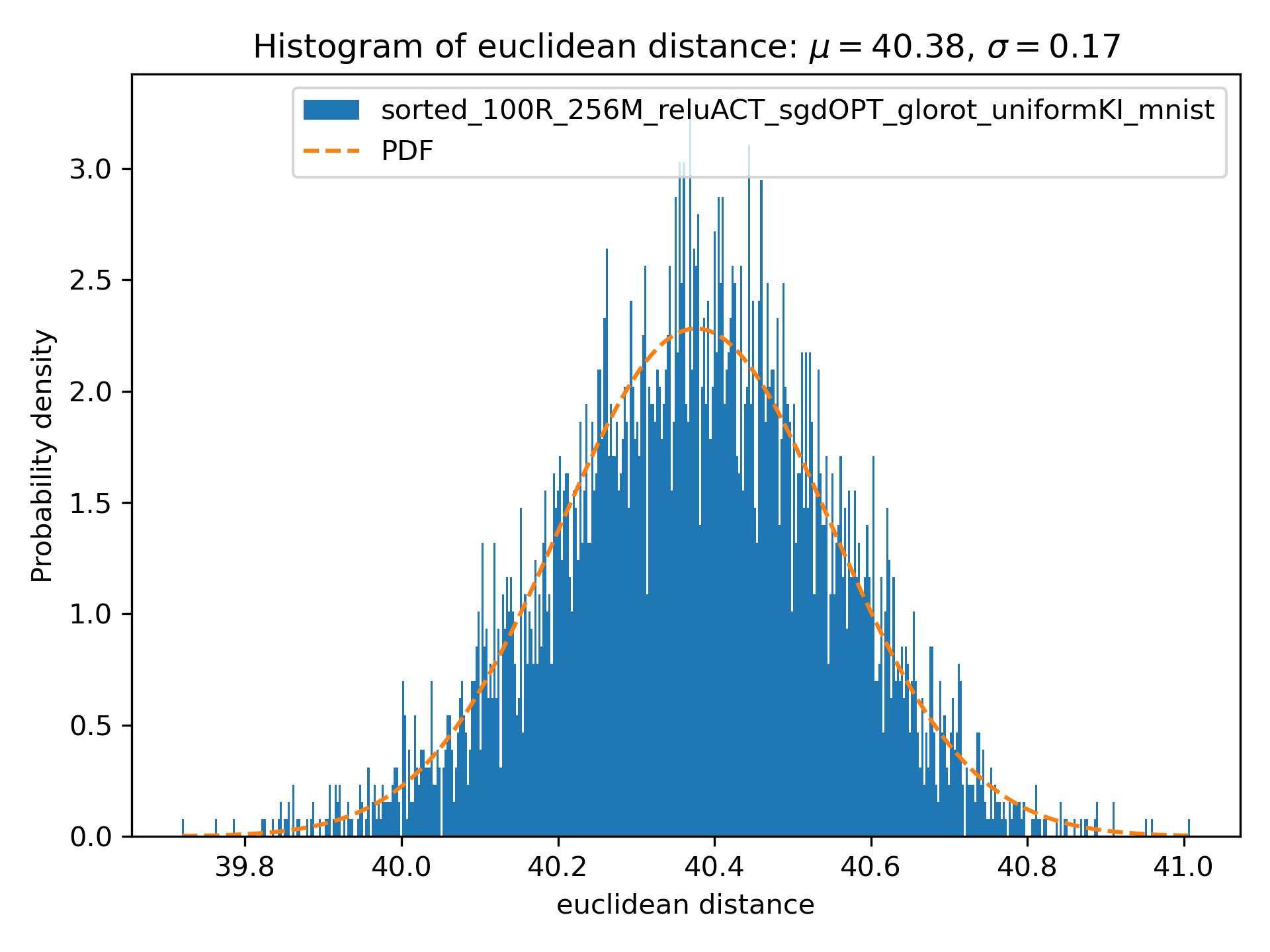}}
	\subfloat[SGD-512, $\mu=45.86$, $\sigma=0.11$]{\includegraphics[width=0.3\linewidth]{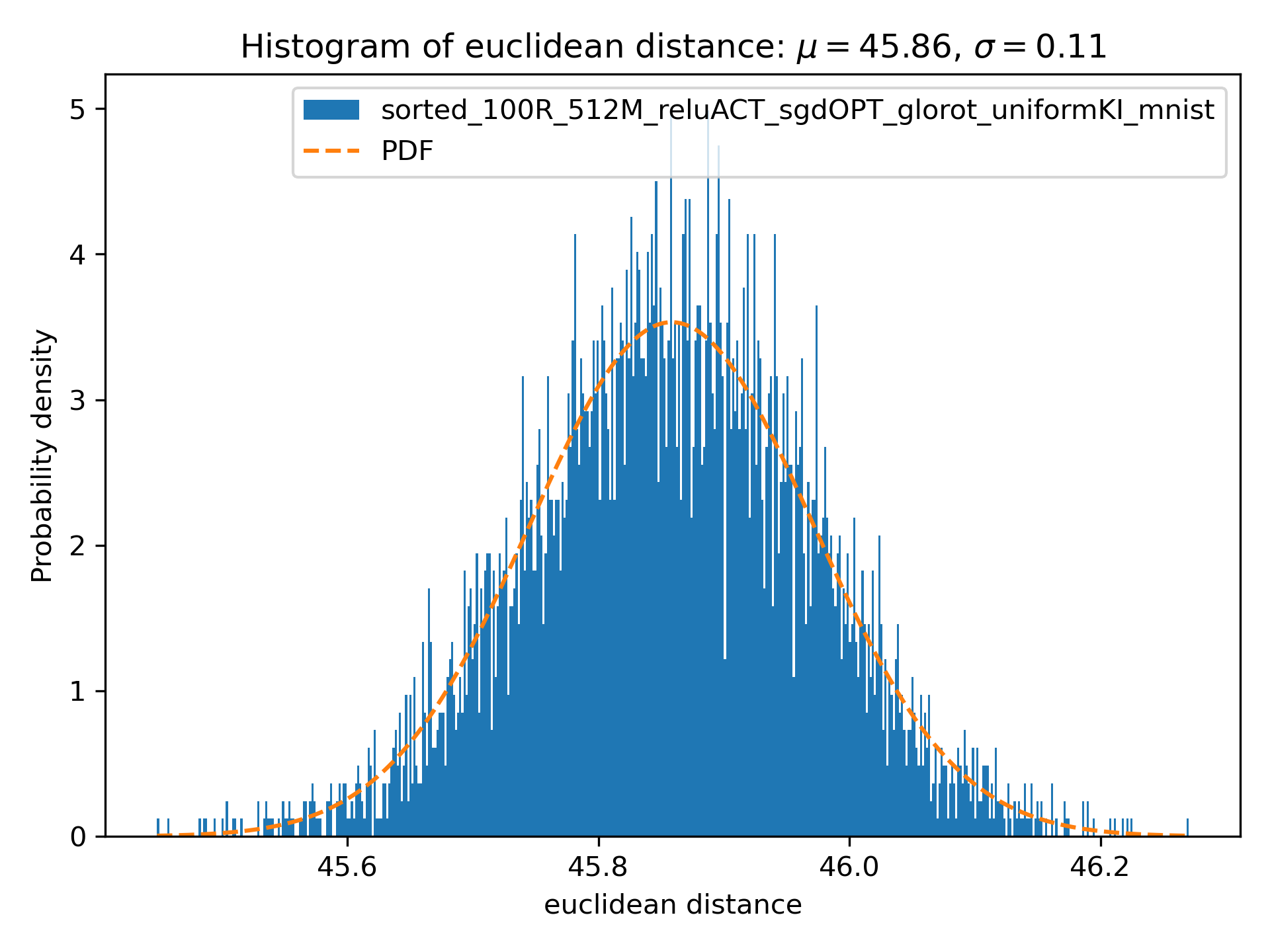}}
        \subfloat[SGD-1024, $\mu=51.44$, $\sigma=0.07$]{\includegraphics[width=0.3\linewidth]{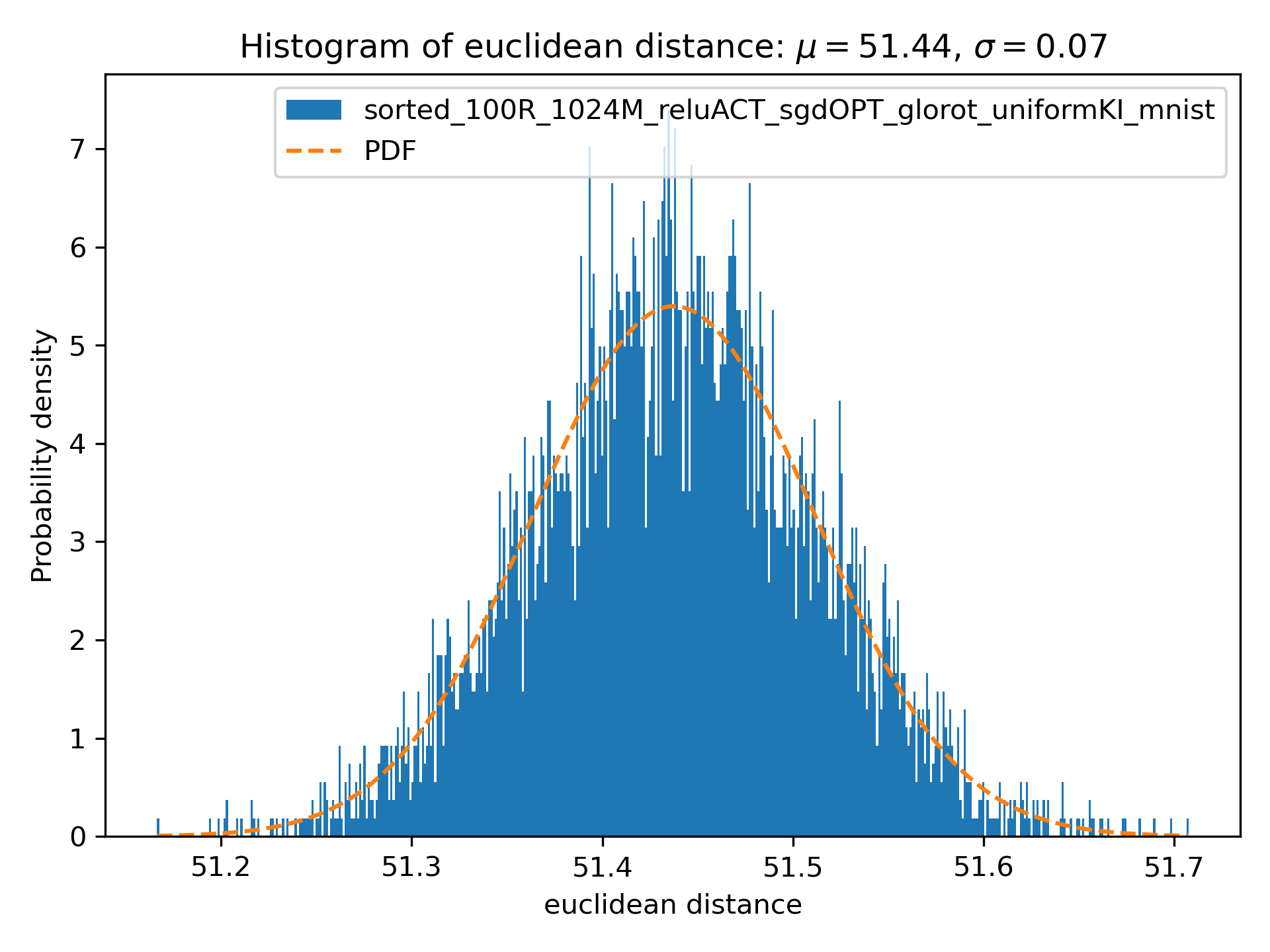}}
\caption{The distribution of Euclidean distance between solution pairs trained for a single hidden layer ReLU model with the Adam optimization algorithm by the MNIST dataset. The training is repeated to obtain $200$ trained solutions. The number before 'M', like '128M', denotes the number of nodes in the hidden layer.}
\label{fig:exp_res4}
\end{figure*}

Here, we analyze a set of different solutions trained for the same DNN by the same dataset, but from different initial conditions. Specifically, we trained a single hidden layer ReLU model on the MNIST dataset using the SGD optimization algorithm, with weights initially randomized from a glorot uniform distribution, repeated for $100$ or $200$ times. These solutions were subsequently reordered using Algorithm \ref{alg:HCA} to mitigate the effect of permutation symmetry in the hidden layers. 
We then compute the Euclidean distance between every pair of the trained solutions as in Eq.(\ref{eq_distance}), and show its distribution in Fig. \ref{fig:exp_res4}. A Gaussian distribution was best-fitted to the distribution for comparison. The figure shows the results for the $FC_1-64$ through $FC_1-1024$ networks, organized from left to right and top to bottom. 

As we can see from Figure \ref{fig:exp_res4}, when the DNNs are trained by the Adam algorithm and the number of nodes in the hidden layer is less than or equal to 256, the distribution fits well to the Gaussian distribution. When the number of nodes in the hidden layer increase to 512 and 1024, the distribution is no longer Gaussian and becomes multi-modal. For instance, 5 peaks are observed for 1024 hidden nodes. These results suggest that solutions are clustered and the clustering is more prominent as the number of hidden nodes increases. We remark that the same clustering is not observed when the DNNs are trained by the SGD algorithm. 

\begin{figure}[t]
    \centering
    \subfloat[Dendrogram-256]{\includegraphics[width=0.5\linewidth]{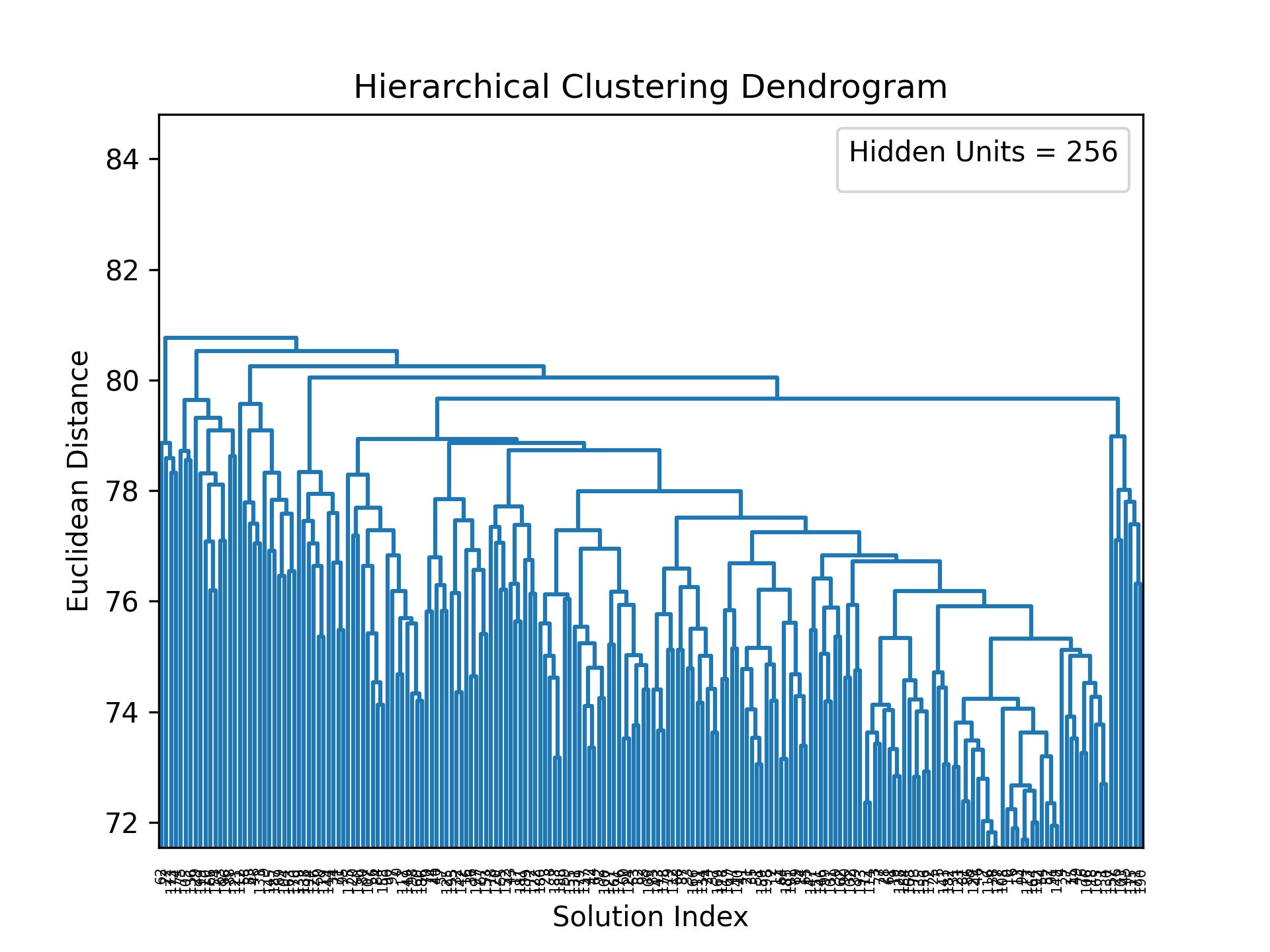}}
    \subfloat[Dendrogram-1024]{\includegraphics[width=0.5\linewidth]{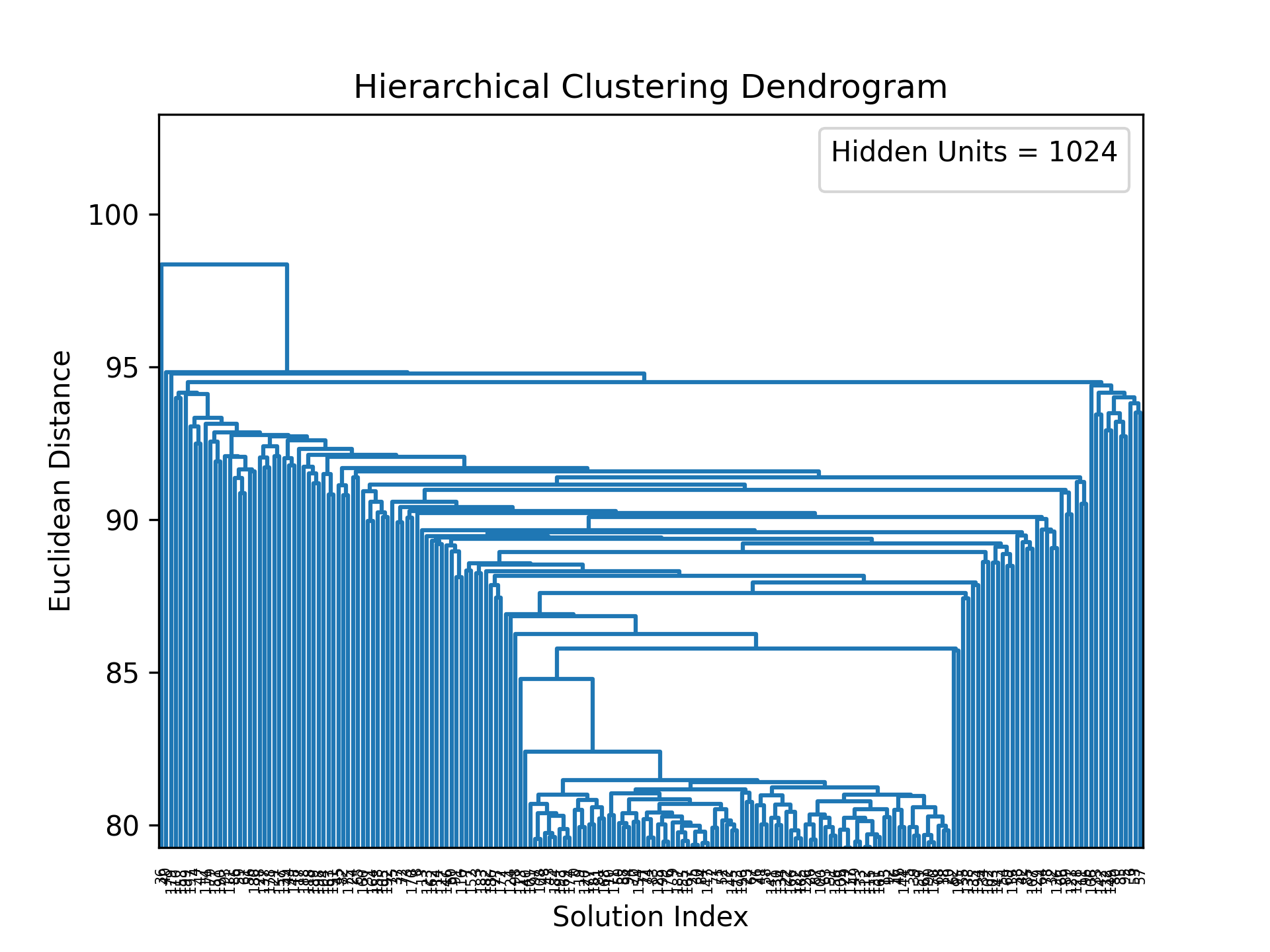}}
    \quad
    \subfloat[Distance matrix-256]{\includegraphics[width=0.5\linewidth]{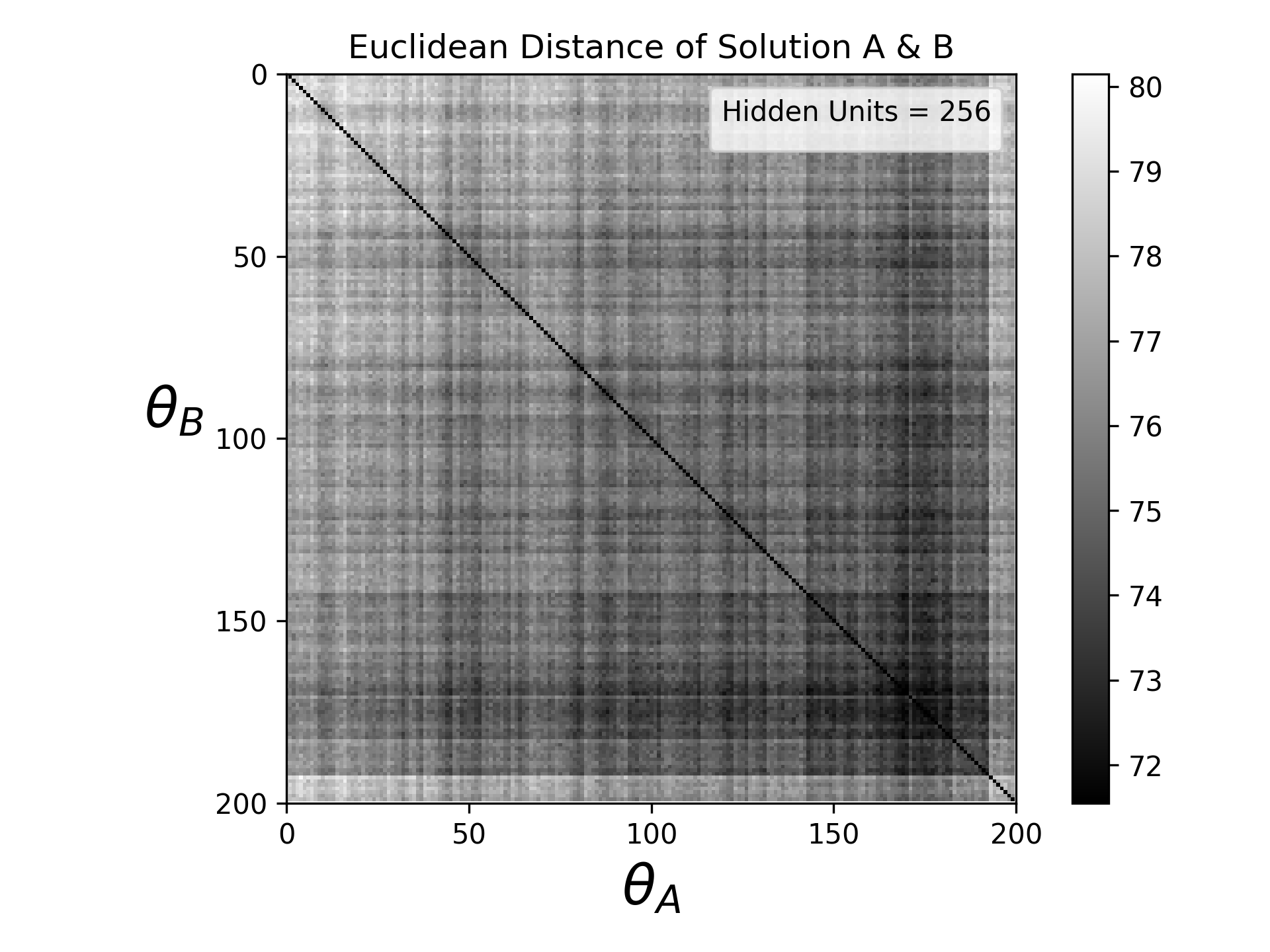}}
    \subfloat[Distance matrix-1024]{\includegraphics[width=0.5\linewidth]{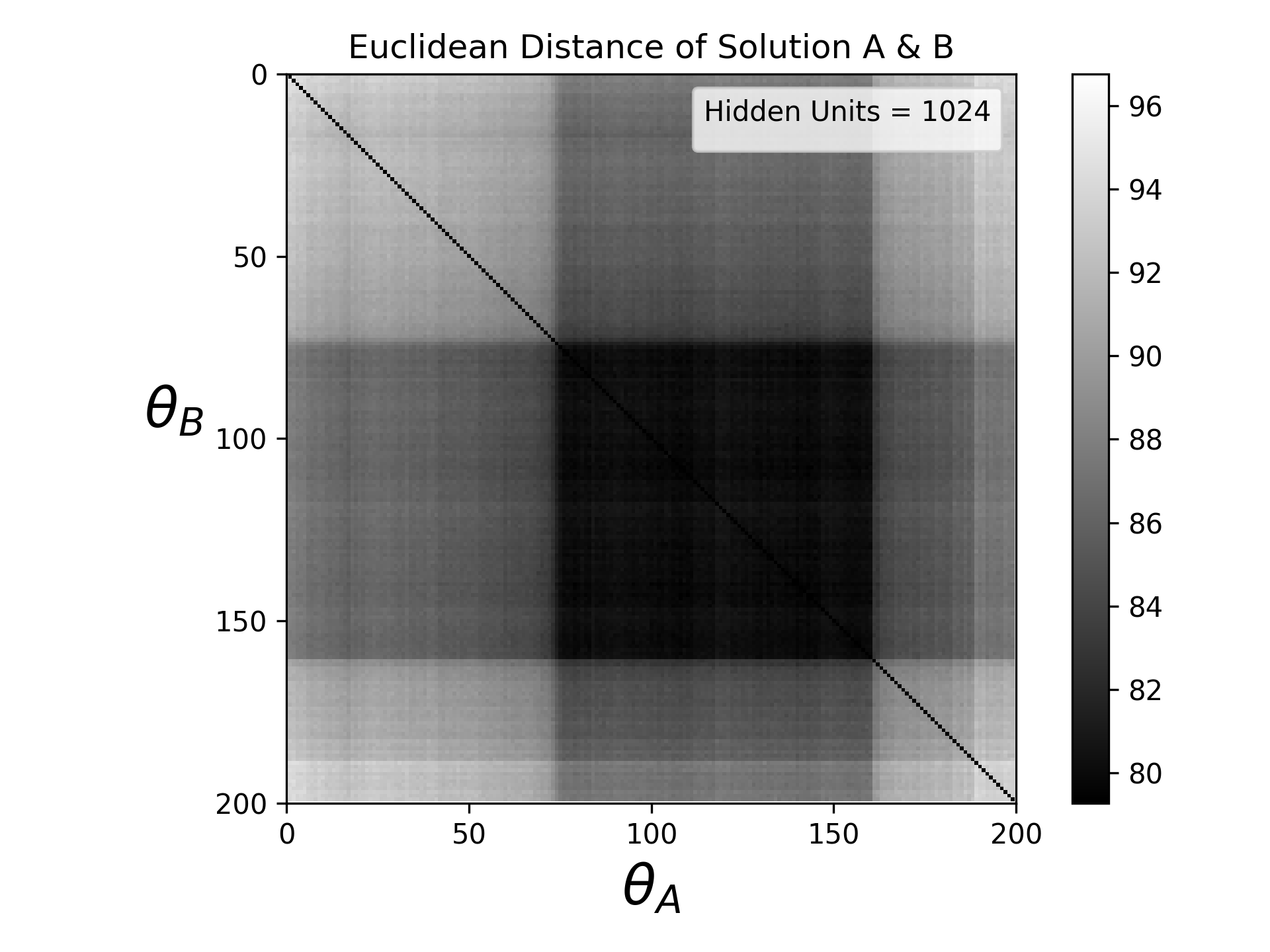}}
    \quad
    \subfloat[Euclidean-256]{\includegraphics[width=0.5\linewidth]{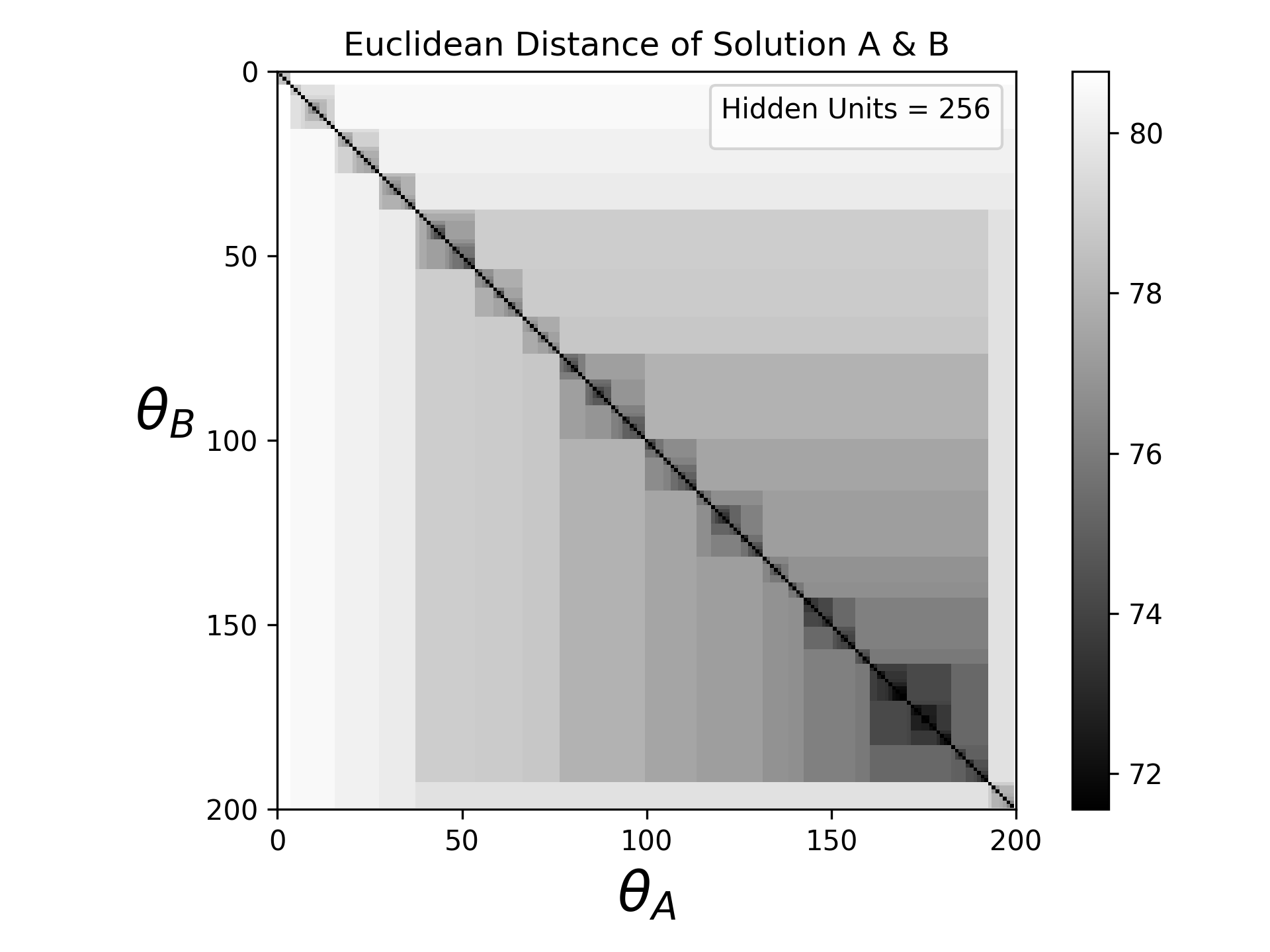}}
    \subfloat[Euclidean-1024]{\includegraphics[width=0.5\linewidth]{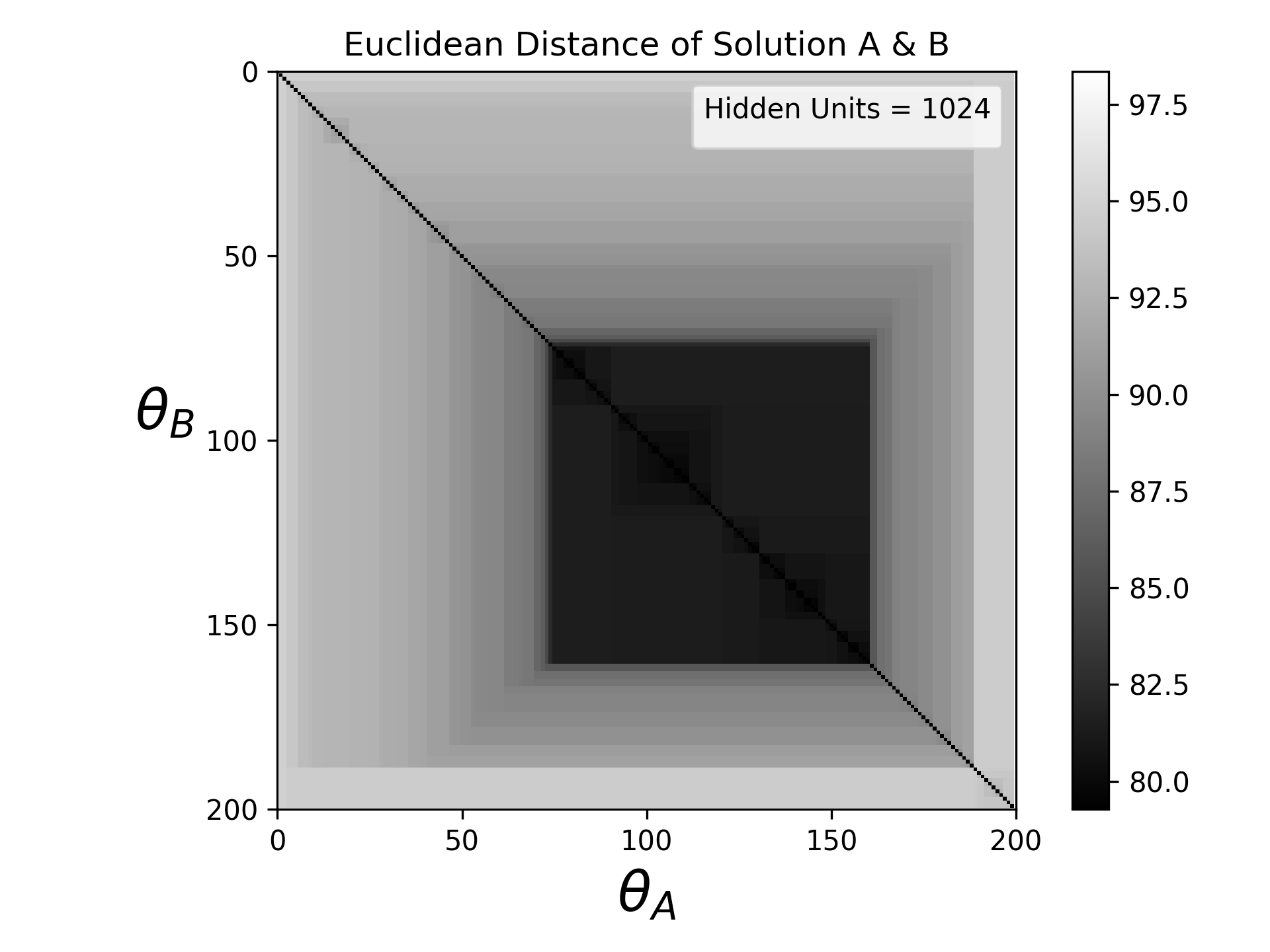}}
    
    \caption{The dendrogram and the similarity matrix. (a) and (b) show dendrograms where leaf nodes represent trained solutions, and intersections indicate cluster formations. The horizontal arrangement reflects their similarity. In (c), (d), (e), and (f), the distance similarity matrices are reordered based on the dendrogram's sequence. Dark colors indicate high similarity (short distances), while light colors indicate low similarity (long distances).}
    \label{fig:exp_res5}
\end{figure}

To further investigate the clustering among the trained solutions, we selected the cases of $FC_1-256$ and $FC_1-1024$ as the representative unimodal and multi-modal distribution for hierarchical clustering of the trained solutions. 
The Dendrogram and the similarity matrix generated by the clustering is shown in Figure \ref{fig:exp_res5}.  

Figure \ref{fig:exp_res5} (a) depicts the hierarchical clustering process of the trained solutions of the $FC_1-256$ network, which contains 256 nodes in the middle layer. And Figure \ref{fig:exp_res5} (c) is the corresponding matrix of Euclidean distance between different pairs of trained solution, where the solutions are reordered according to the results of hierarchical clustering. 
The gray square has no obvious hierarchical structure and is consistent with the characteristics of the tree diagram. 
Figure \ref{fig:exp_res5} (e) is the matrix showing the cluster distance, and multiple clear blocks on the diagonal can be observed.
Figure \ref{fig:exp_res5} (b), (d) and (f) are similar to (a), (c) and (e), but for the $FC_1-1024$ network, which contains 1024 nodes in the middle layer. We can see in Figure \ref{fig:exp_res5} (d) a dark black square near the main diagonal. The clustered structure is obvious in this case. In Figure \ref{fig:exp_res5} (f), which is based on the cluster distance, there are also multiple blocks. These results share similarity with the grouping of spin configurations in the RSB phase as depicted in Figure \ref{fig:rsb}. There are multiple local minima in the loss landscape of a single-hidden-layer ReLU network and when the number of parameters in the network meets certain conditions, the Adam optimization algorithm can be used to train and obtain solutions located in different local minima.

\subsection{Extreme value flatness analysis in loss landscape}

The stochastic gradient descent (SGD) method and its variants are the algorithm of choice for many deep learning tasks. These methods operate in small-batch mode, where a portion of the training data is sampled to calculate an approximation of the gradient. It has been observed in practice that the quality of the model, as measured by its generalization ability, decreases when using larger batches.

Small mini-Batch (SB) denotes small batch training, which is fixed at 256, and Large mini-batch (LB) denotes large batch training, which is set to 10\% of the samples in the training data set. 
Adam optimization algorithm is used to train the neural network model, and two sets of solutions $\theta_{SB}$ and $\theta_{LB}$ are obtained respectively. 
All experiments were performed $5$ times from different starting points, and the mean and standard deviation were recorded. Then use the linear interpolation algorithm for any pair of $\theta_{SB}$ and $\theta_{LB}$:

\begin{equation}
    \theta = (1-a)*\theta_{SB} + a*\theta_{LB},~a\in[-1,2]
\end{equation}

\begin{figure}[t]
    \centering
    \includegraphics[width=\columnwidth]{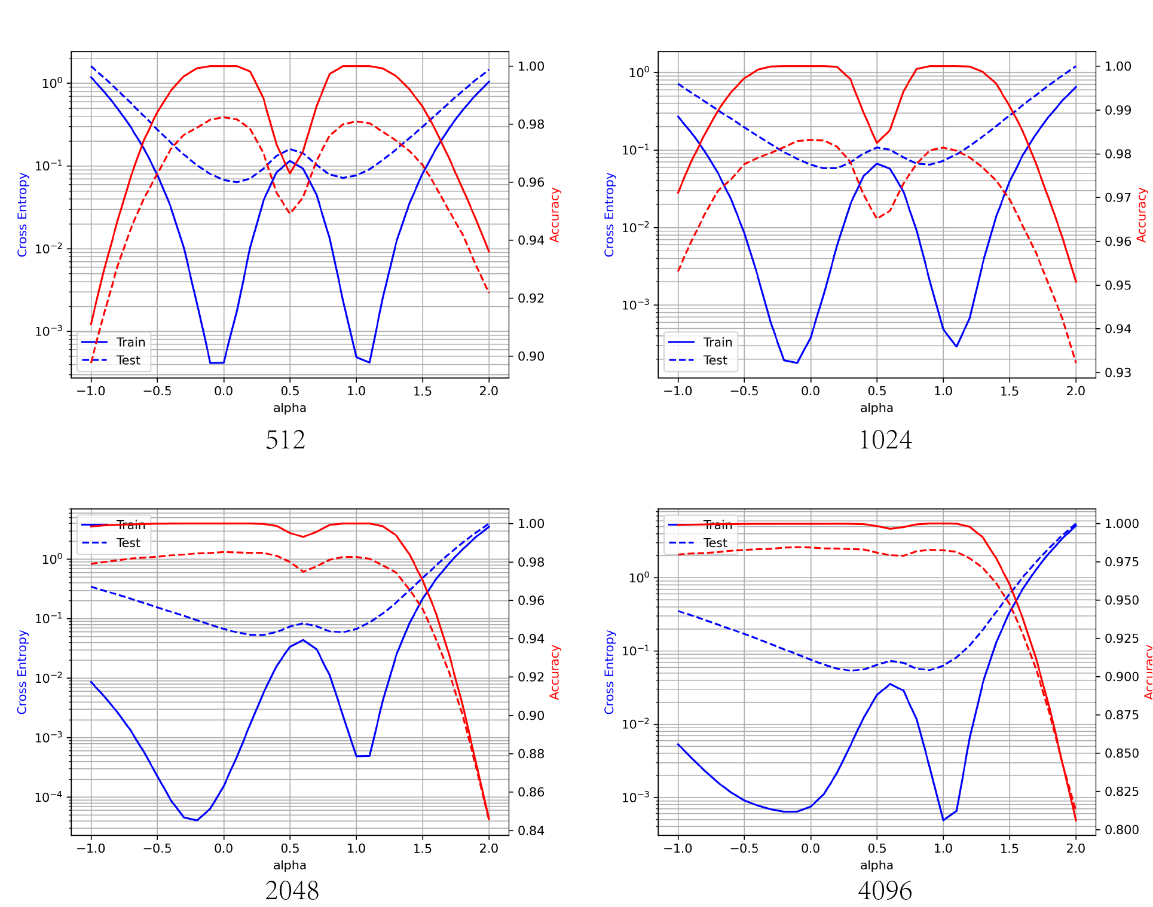}
    \caption{The loss function and accuracy for the training and test data sets. The number below each subplot, i.e. 512, 1024, 2048, 4096, represents the number of nodes in the middle layer. The solid line represents the training data set, and the dotted line represents the testing data set. The vertical axis for the blue lines is 'Cross Entrophy Loss'. And the vertical axis for the red lines is 'Accuracy'.}
    \label{fig:exp_res8}
\end{figure}

Figure \ref{fig:exp_res8} shows the loss function and accuracy of the training and test data sets in the loss landscape along the path containing two solutions. The solid line represents the training data set, the dotted line represents the test data set.
$\alpha=0$ represents the minimum value of the small batch, and $\alpha=1$ represents a large batch minimum. 
It can be seen that the line is significantly flatter at the minimum value of the small batch than at the minimum value of the large batch, and as the number of hidden layer nodes increases, the minimum value of the small batch becomes flatter. 
Besides, the test accuracy at $\alpha=0$ is higher than at $\alpha=1$. 
Therefore, it can be concluded that flat minima have better generalization performance.

Please note that additional experiments are detailed in Supplementary Materials, where further analyses and supplementary results are provided to support the findings discussed above.

\section{Conclusion} 

Although deep learning has many successful application cases in various fields, understanding of its underline mechanism remains limited. To this end, this paper delves into the loss landscape of single hidden layer ReLU neural network by drawing analogy with energy landscape in spin glass in statistical physics. Experimental results demonstrated the existence of multiple extrema in the loss landscape, and empirically established that improved generalizability is correlated with the flatter local minima in loss landscape.
However, this study's focus on the most basic model of DNNs, which limits the generalizability of our findings to more complex or different network architectures. Furthermore, the empirical nature of our results and the complex methodologies employed, such as the MCMC random walk and permutation-interpolation protocols, may pose challenges in reproducibility and practical applications. These techniques, while innovative, require a deep understanding, which could hinder their adoption in broader research context.
In conclusion, by shedding light on the underpinnings of the loss landscape in deep neural networks, this work contributes to a deeper understanding of DNNs through a new perspective and a new set of tools.

\begin{ack}

CHY fully supported by the Research Grants Council of the Hong Kong Special Administrative Region, China (Projects No. GRF 18304316, GRF 18301217, GRF 18301119 and GRF 18300623), the Dean's Research Fund of the Faculty of Liberal Arts and Social Sciences, The Education University of Hong Kong, Hong Kong Special Administrative Region, China (Projects No: FLASS/DRF 04418, FLASS/ROP 04396 and FLASS/DRF 04624), and the Research Development Office Internal Research Grant, The Education University of Hong Kong, Hong Kong Special Administrative Region, China (Projects No. RG67 2018-2019R R4015 and No. RG31 2020-2021R R4152).

\end{ack}

\bibliographystyle{unsrt}
\bibliography{main}

\newpage
\appendix

\section{Appendix}

\setcounter{table}{0}
\setcounter{figure}{0}
\setcounter{algorithm}{0}
\renewcommand{\thetable}{A\arabic{table}}
\renewcommand{\thefigure}{A\arabic{figure}}
\renewcommand{\thealgorithm}{A\arabic{algorithm}}

\subsection{Information of Datasets}\label{app:datasets}

\begin{itemize}
    \item The {\bf MNIST} data set is an image classification data set commonly used for machine learning. It is a dataset of handwritten digit images and has more than $60,000$ training images and more than $10,000$ test images, each of which is a $28\times 28$ pixel grayscale image representing a number between $0$ and $9$. These numbers are dimensionally normalized and centered within a fixed-size image with pixel values between $0$ and $255$, representing the brightness of the image.

    \item The {\bf Fashion-MNIST} is an image classification dataset that contains $60,000$ training images and $10,000$ test images, each image is a $28\times 28$ grayscale image. Its content includes 10 categories of clothing, namely: T-shirt/top, Trouser, Pullover, Dress, Coat, Sandal, Shirt, Sneaker, Bag, Ankle boot.

    \item The {\bf CIFAR-10} dataset is a commonly used dataset for image classification and contains $50,000$ training images and $10,000$ test images. Each image is a color image with the size of $32\times 32\times 3$. There are $10$ categories of images, namely: cars, birds, cats, frogs, deer, dogs, airplanes, trucks, boats and horses. 

    \item {\bf CIFAR-100} is a new and more comprehensive image classification dataset, which contains $50,000$ training images and $10,000$ test images. Each image is a color image with the size of $32\times32\times3$. Compared with the CIFAR-10 dataset, the CIFAR-100 dataset contains more than $100$ different categories. The 100 categories in the CIFAR-100 dataset are divided into $20$ supercategories and $100$ subcategories, with each supercategory contains 5 subcategories.
\end{itemize}

\begin{table}[t]
    \small
    \centering
    
    \begin{tabular}{@{}l@{}ccc@{}c@{}}
         \toprule
         \textbf{Dataset} 
         & \textbf{\begin{tabular}[c]{@{}l@{}} Training \\ Set\end{tabular} } 
         & \textbf{\begin{tabular}[c]{@{}l@{}} Testing \\ Set\end{tabular} }  
         & \textbf{\begin{tabular}[c]{@{}l@{}} Feature \\ Size\end{tabular} } 
         & \textbf{Category} \\ \hline
         MNIST & $60{,}000$ & $10{,}000$ & $28\times 28$ & 10 \\ 
         Fashion-MNIST & $60{,}000$ & $10{,}000$ & $28\times 28$ & 10 \\ 
         CIFAR-10 & $50{,}000$ & $10{,}000$ & $32\times 32\times 3$ & 10 \\
         CIFAR-100 & $50{,}000$ & $10{,}000$ & $32\times 32\times 3$ & 100 \\
    \bottomrule
    \end{tabular}
    \caption{General Statistical Information of the Datasets}
    \label{tab:dataset}
\end{table}

\subsection{Neural Network parameters}\label{app:nnp}

The Deep Neural Networks (DNNs) utilized in this study are structured with three layers. The DNNs are categorized into two sets based on different configurations of the input layers. Specifically, the input layer dimensions for $FC_1$ and $FC_2$ are $784$ and $3,072$, respectively. The hidden layers consist of $64$, $128$, $256$, $512$, and $1024$ nodes. The ReLU function is employed to activate the hidden layers, while the output layer, containing $10$ neurons, is activated by the Softmax function.

\begin{table}[t]
    \small
    \centering
    
    \begin{tabular}{lccc}
        \toprule
        \textbf{Name} & \textbf{Parameters} 
        & \textbf{\begin{tabular}[c]{@{}l@{}} Input \\ Layer \end{tabular}} 
        & \textbf{\begin{tabular}[c]{@{}l@{}} Hidden \\ Layers \end{tabular}} \\ 
        \midrule
        $FC_1$-64 & $50{,}890$ & \multirow{5}{*}{$784$} & $64$ \\
        $FC_1$-128 & $103{,}178$ &  & $128$  \\
        $FC_1$-256 & $203{,}530$ &  & $256$  \\
        $FC_1$-512 & $407{,}050$ &  & $512$  \\
        $FC_1$-1024 & $814{,}090$ &  & $1,024$ \\
        \midrule
        $FC_2$-64 & $197{,}322$ & \multirow{5}{*}{$3,072$} & 64 \\
        $FC_2$-128 & $394{,}634$ &  & $128$ \\
        $FC_2$-256 & $789{,}258$ &  & $256$ \\
        $FC_2$-512 & $1{,}578{,}506$ &  & $512$ \\
        $FC_2$-1024 & $3{,}157{,}002$ &  & $1,024$ \\
        \bottomrule
    \end{tabular}
    \caption{Neural Network Setup}
    \label{tab:neural_network}
\end{table}

\subsection{Theorems and Proofs}\label{app:theorems}

The set of all weights in a permutation network is denoted as the weight vector $\vec{\textbf{w}}$, with $\vec{\textbf{w}}\in \mathbb{R}^q$, where $q$ is the total number of real-valued adjustable parameters. The set of all permutations of $\vec{\textbf{w}}$ forms the non-Abelian group $S_q$ (the $q$-th symmetry group), referred to as $S$. This group contains $q!$ elements. For a permutation neural network, $T$ is defined as the set of all non-singular linear weight transformations (e.g., reversible linear mappings from $\mathbb{R}^q$ to $\mathbb{R}^q$ that preserve the network's input/output transformation function.
By definition, $T$ must include the weights of pairs of hidden layer units in the permutation network but can also encompass additional transformations. For instance, in a backpropagation network, reversing the signs of all weights of a hidden layer unit, reversing the signs of the input weights associated with that unit's output in the next layer, and adjusting their bias weights by the initial values of these input weights is a transformation that leaves the network's input/output functions unchanged. This transformation is not a permutation but is included in $T$.
Finally, $T$ for any permutation neural network is a subset of the general linear group $GL(q,R)$, which comprises all reversible linear transformations from $\mathbb{R}^q$ to $\mathbb{R}^q$. Based on these definitions, Theorem \ref{thm:rpl} can be established:

\begin{theorem}
If $T$ is a proper subgroup of $GL(q,R)$, then the number of elements in $T$ satisfies $\#T \geq\prod_{i=1}^K M_i!$ , where $M_i$ represents the number of neural units in the $i$-th hidden layer, and $K$ represents the number of hidden layers.

\end{theorem}

\begin{proof}
To prove that $T$ is a subgroup, it is necessary to prove that $T$ satisfies closure property under identity elements, inverse elements and multiplication operations. Since transformation of any element of $T$ leaves the network's input/output transformation function unchanged, the product operation is also in $T$. $T$ satisfies closure under the inverse element operation because the inverse element of a weight transformation in $T$ simply restores the network to its pre-transformation state, thus retaining the same overall network input/output transformation function. Therefore, $T$ is a subgroup of $GL(q,R)$. In fact, $T$ is a proper subgroup, for example, the permutation of the two weights of a single first hidden layer unit (which is a reversible linear transformation in $S$, itself a proper subgroup of $GL(q,R))$ cannot be determined by $T$ composed of any elements. Permutation of two weights for a single first hidden layer unit cannot be composed of any elements in $T$.

A lower bound on the number of elements in $T$ can be determined by considering the hidden layers in order from the first (the layer that receives input from the input unit) to the last. Any permutation of units in the first hidden layer can be generated by permutations between pairs of units in that layer. Thus, on the first hidden layer, $T$ contains the group of all permutations of the first hidden layer units where the weight ordering remains constant within the units. The number of elements in this group is $M_1!$ .

The weight exchange arrangement of the second hidden layer unit is isomorphic to the group of all second hidden layer unit arrangements. The number of elements in this group is $M_2!$. This process is repeated for each hidden layer until the last hidden layer $K$. Since the arrangement of units selected in each hidden layer is independent of the arrangement of units selected in the previous hidden layer, there will be a total of $\prod_{i=1}^K M_i!$ in $T!$. Typically, $T$ will be larger than this because if we first perform the permutation on the second hidden layer and then the first hidden layer, we will end up with a different permutation than the one found by the previous process. Additionally, non-permutational transformations may also occur in $T$. Therefore, the number of elements in $T$ satisfies $\#T \geq\prod_{i=1}^K M_i!$.

\end{proof}

\begin{theorem}\label{thm:similarity}
There is a connection between the highly non-convex loss function of the fully connected feedforward neural network model and the Hamiltonian of the spherical spin glass model under the following assumptions: i) variable independence, ii) network parameter redundancy, iii) uniformity.
\end{theorem}

\begin{proof}
For theoretical analysis, we consider a simple model of a fully connected feedforward deep network with a single output and ReLU activation, called network $\mathcal{N}$. 
Suppose we focus on binary classification tasks. Let $X$ be a random input vector of dimension $d$. Let $(H-1)$represent the number of hidden layers in the network, the input layer is the $0$-th layer, and the output layer is the $H$-th layer. Let $n_i$ represent the number of nodes in the $i$-th layer ($n_0=d$,$n_H=1$). 
Let $W_i$ be the weight matrix between the $(i-1)$-th layer and the $i$-th layer in the network. Furthermore, let $\sigma$ represent the activation function, $\sigma(x)=max(0,x)$. Therefore, the network output expression is:

\begin{equation}
    Y=q\sigma(W_H^T\sigma(W_{H-1}^T\ldots\sigma(W_1^TX))\ldots)
\end{equation}

where $q=\sqrt{(n_0n_1\ldots n_H)^{(H-1)/2H}}$ is just a normalization factor. In order to make the output more similar to the output in the spin glass model, assuming that there are $\gamma$ different roads from a certain input node to the output node, $\gamma=(n_1n_2\ldots n_H)$, then the network output can also be re-expressed in the following way:

\begin{align}
    Y&=q\sum_{i=1}^{n_0}X_i\sum_{j=1}^{\gamma}A_{i,j}\prod_{k=1}^H w_{i,j}^{(k)} \nonumber \\
    &=q\sum_{i=1}^{n_0}\sum_{j=1}^{\gamma}X_{i,j}A_{i,j}\prod_{k=1}^H w_{i,j}^{(k)}
\end{align}

where $A_{i,j}$ represents whether the $j$-th path of the $i$-th input node is activated. Let $\Psi=\prod_{i=0}^H n_i=n_0\gamma$ be the total number of paths in the model, $\Lambda=\sqrt[H]{\Psi}$, $N=\sum_{i=0}^{H-1}n_in_{i+1}$ is the total number of parameters.

(1) variable independence

For the input $X_{i,j}$, it is assumed that their distribution is standard normal and the variables are independent of each other.
Path: Assume that the probability of each path activation is $\rho$, that is, $\{A_{i,j}\}~i.i.d$  is a Bernoulli distribution with a success probability of $\rho$. For the weights, number them from $1$ to $N$, record the weight of a certain path passing layer by layer as $(w_{i_1},...,w_{i_H})$, $r_{i_1,i_2,\ldots,i_H}$ indicates whether there is a path passing through these The path of the weight, therefore, $\sum_{i_1, i_2,\ldots,i_H=1}^Nr_{i_i,i_2,\ldots,i_H}=\Psi$, therefore,

\begin{equation}
    Y_N\coloneqq\mathbb{E}_A[Y]
\end{equation}

\begin{equation}
    \mathbb{E}_A[Y]=q\sum_{i_1,i_2,\ldots,i_H=1}^N\sum_{j=1}^{r_{i_1,i_2,\cdots,i_H}}X_{i_1,i_2,\ldots,i_H}^{(j)}\rho\prod_{k=1}^H w_{i_k}
\end{equation}

(2) network parameter redundancy

The vast majority of weights (even as high as 95\%) are redundant. After deleting them (fixed to 0 and not updated), the accuracy on the test set remains basically unchanged. Suppose this view is correct.

(3) uniformity

It is not redundant to assume that network $\mathcal{N}$ has only $s$ weights (assume it is the first $s$), assuming that they are close to evenly distributed on this network structure. Note that the network after deleting excess hits is $\mathcal{M}$, and the corresponding output of the above formula in this network is $Y_s$. If the difference in prediction accuracy between the two networks does not exceed $\epsilon$, then $\mathcal{M}$ is called a $(s,\epsilon)$-reduction image of $\mathcal{N}$.

\begin{equation}
    Y_s=q\sum_{i_1,i_2,\ldots,i_H=1}^s\sum_{j=1}^{t_{i_1,i_2,\ldots,i_H}}X_{i_1,i_2,\ldots,i_H}^{(j)}\rho\prod_{k=1}^Hw_{i_k}
\end{equation}

Among them, $t_(i_1,i_2,\ldots,i_H)$ represents the number of times $(w_{i_1},\ldots,w_{i_H})$ appears in the above formula. If it is uniformly distributed, then the probability of each path chooses a certain weight from $(w_{i_1},\ldots,w_{i_H})$ is $1/s^H$, so $t_{i_1,i_2,\ldots,i_H}=\Psi/{s^H}$. Based on the above assumptions, there exists a constant $c$ such that:

\begin{equation}
    \frac{1}{c}\cdot\frac{\Psi}{s^H}\leq t_{i_1,i_2,\ldots,i_H}\leq c\cdot\frac{\Psi}{s^H}
\end{equation}

In light of this, consider

\begin{align}
    \hat{Y}&=q\sum_{i_1,i_2,\ldots,i_H=1}^s\sum_{j=1}^{\psi/{s^H}}X_{i_1,i_2,\ldots,i_H}^{(j)}\rho\prod_{k=1}^H w_{ik} \nonumber \\
    &=q\sum_{i_1,i_2,\ldots,i_H=1}^{\Lambda}X_{i_1,i_2,\ldots,i_H}\rho\prod_{k=1}^Hw_{ik} 
\end{align}

Note that $\Psi/{s^H}=(\Lambda/s)^H$. Therefore, we can obtain the second equation by recoding.
Assume that the weights after recoding satisfy the spherical constraint:

\begin{equation}
    \frac{1}{\Lambda}\sum_{i=1}^{\Lambda}w_i^2=C
\end{equation}

Based on the above assumptions, the loss function in the neural network can be regarded as the Hamiltonian in the spin glass model. First, the Hamiltonian of the $H$-spin spherical spin glass model can be written as:

\begin{equation}
    \mathcal{L}_{\Lambda,H}(\tilde{\textbf{w}})=\frac{1}{\Lambda^{\frac{(H-1)}{2}}}\sum_{i_1, i_2, \ldots, i_H=1}^{\Lambda}X_{i_1,i_2,\ldots,i_H}\tilde{w}_{i_1}\tilde{w}_{i_2}\ldots\tilde{w}_{i_H}
\end{equation}

The sphere constraint is $\frac{1}{\Lambda}\sum_{i=1}^{\Lambda}\tilde{w}_i^2=1$.

The absolute loss and hinge loss of Neural Networks are as follows:

\begin{equation}
    \mathcal{L}_{\Lambda,H}^a(\textbf{w})=\mathbb{E}_A\left[\vert Y_t-Y\vert\right]
\end{equation}

\begin{equation}
    \mathcal{L}_{\Lambda,H}^h(\textbf{w})=\mathbb{E}_A\left[\mathrm{max}(0, 1-Y_tY)\right]
\end{equation}

where $Y_t$ is a random variable with respect to the real data labels. If you use the previous $\hat{Y}$ to estimate $Y_N=\mathbb{E}_A[Y]$, both losses can be written in the following form:

\begin{equation}
    \mathcal{L}_{\Lambda, H}(\tilde{\textbf{w}})=C_1+C_2q\sum_{i_1,i_2,\ldots,i_H=1}X_{i_1,i_2,\ldots,i_H}\prod_{k=1}^H\tilde{w}_{i_k}
\end{equation}

\end{proof}

Neural networks are often analyzed using tools developed in spin glass research because both systems are optimization problems with quenched constraints. 
Through theoretical analysis, we establish the connection between the highly non-convex loss function of the DNN model and the Hamiltonian of the spin glass model. 
We demonstrate the similarity between the loss function in DNNs and the Hamiltonian in spin glass model by Theorem \ref{thm:similarity}.
Based on this, we can transfer statistical physics methods for analyzing spin glass into the analysis of the loss landscape of DNNs.

\subsection{Additional experimental results}\label{app:exp_res}

\subsubsection{Random Walking in Parameter Space}

\begin{figure}
    \centering
    \includegraphics[width=\columnwidth]{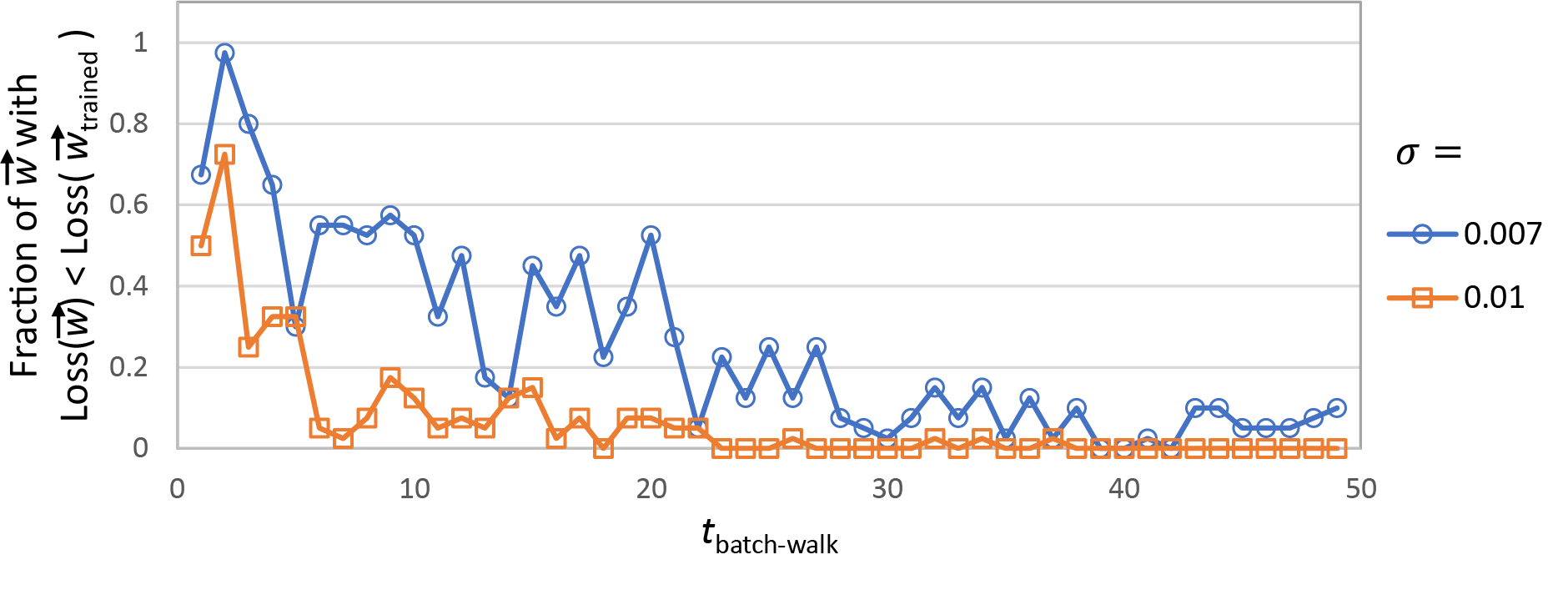}
    \caption{The experimental results on random walks in the parameter configurational space from the trained parameter configurations $\vec{\textbf{w}}_{trained}$. The random walk only passes through $\vec{\textbf{w}}$ with loss less than the original trained loss $L(\vec{\textbf{w}}_{trained})$.}
    \label{fig:random_walk_cmp}
\end{figure}

The result shown in Figure \ref{fig:random_walk_cmp} is obtained by only walking paths with loss $L[\vec{\textbf{w}}(t)]$ less than $L[\vec{\textbf{w}}_{trained}]$. 
In each time step $t_{batch-walk}$, we conduct 40 individual random
walk trials staring from the previous $\vec{\textbf{w}}(t_{batch-walk}-1)$, and compute the fraction that among the 40 newly drawn $\vec{\textbf{w}}$, their loss is less than the original trained loss $L[\vec{\textbf{w}}_{trained}]$. As we can see, this fraction decreases as $t_{batch-walk}$ increases, which may imply that the random walk trajectories are approaching the boundary of the attraction basins: as $t_{batch-walk}$ increases, less neighboring $\vec{\textbf{w}}$ of $\vec{\textbf{w}}(t_{batch-walk})$ have a loss less than the original trained loss. The set-up and the training of the neural network are identical to those in the experiment of Figure \ref{fig:random_walk_cmp}, where the different $\sigma$ of the two cases characterize the magnitude of parameter $\Delta\vec{\textbf{w}}$ changes drawn from the distribution $P(\Delta\vec{\textbf{w}})=\mathcal{N}(0, \sigma^2)$. 

\subsubsection{Comparison of Different Optimization Methods}

\begin{figure*}[t]
	\small
	\centering
        \subfloat[Adam-64, $\mu=32.64$, $\sigma=0.42$]{\includegraphics[width=0.3\linewidth]{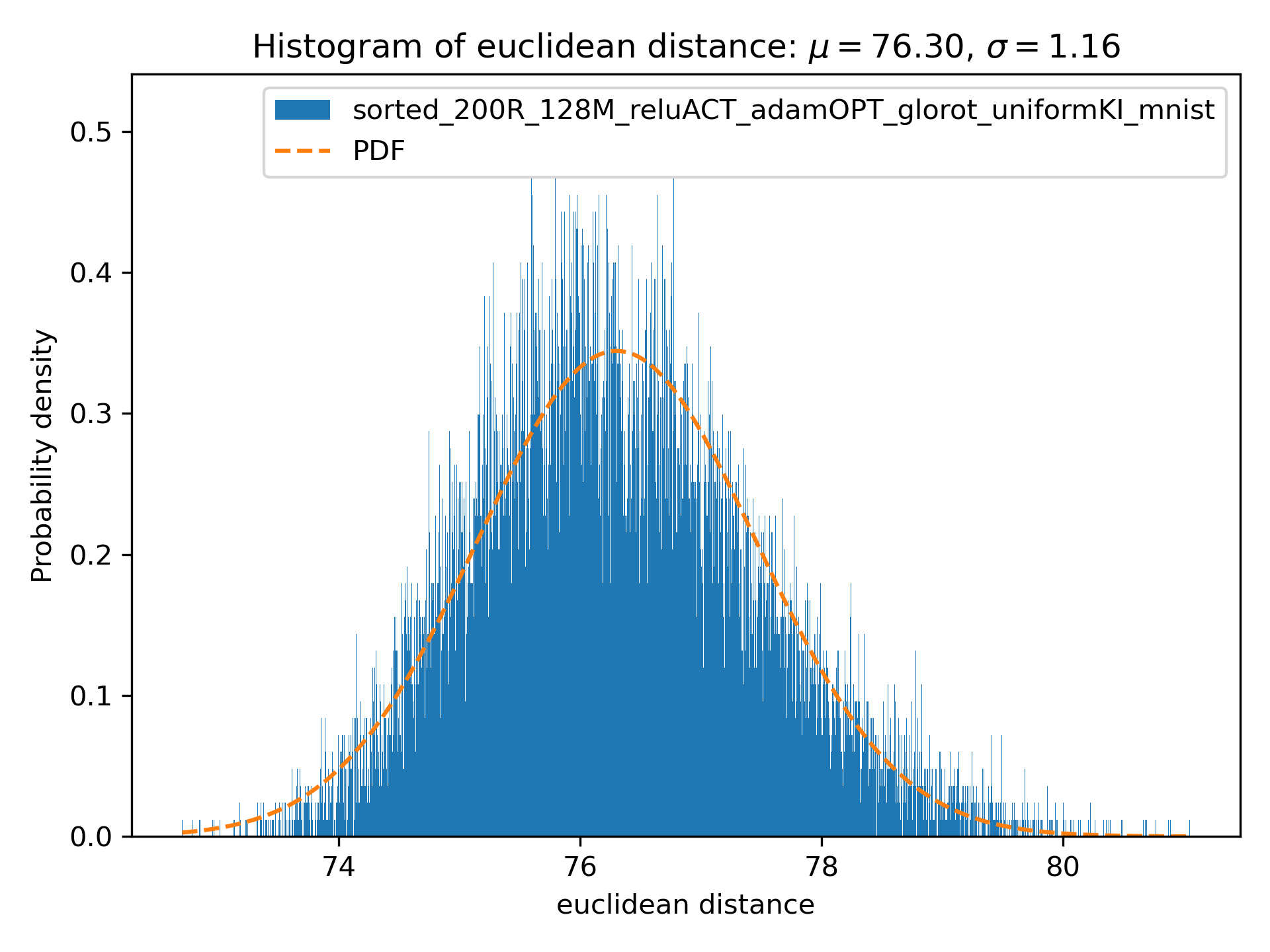}}
        \subfloat[Adam-128, $\mu=35.90$, $\sigma=0.27$]{\includegraphics[width=0.3\linewidth]{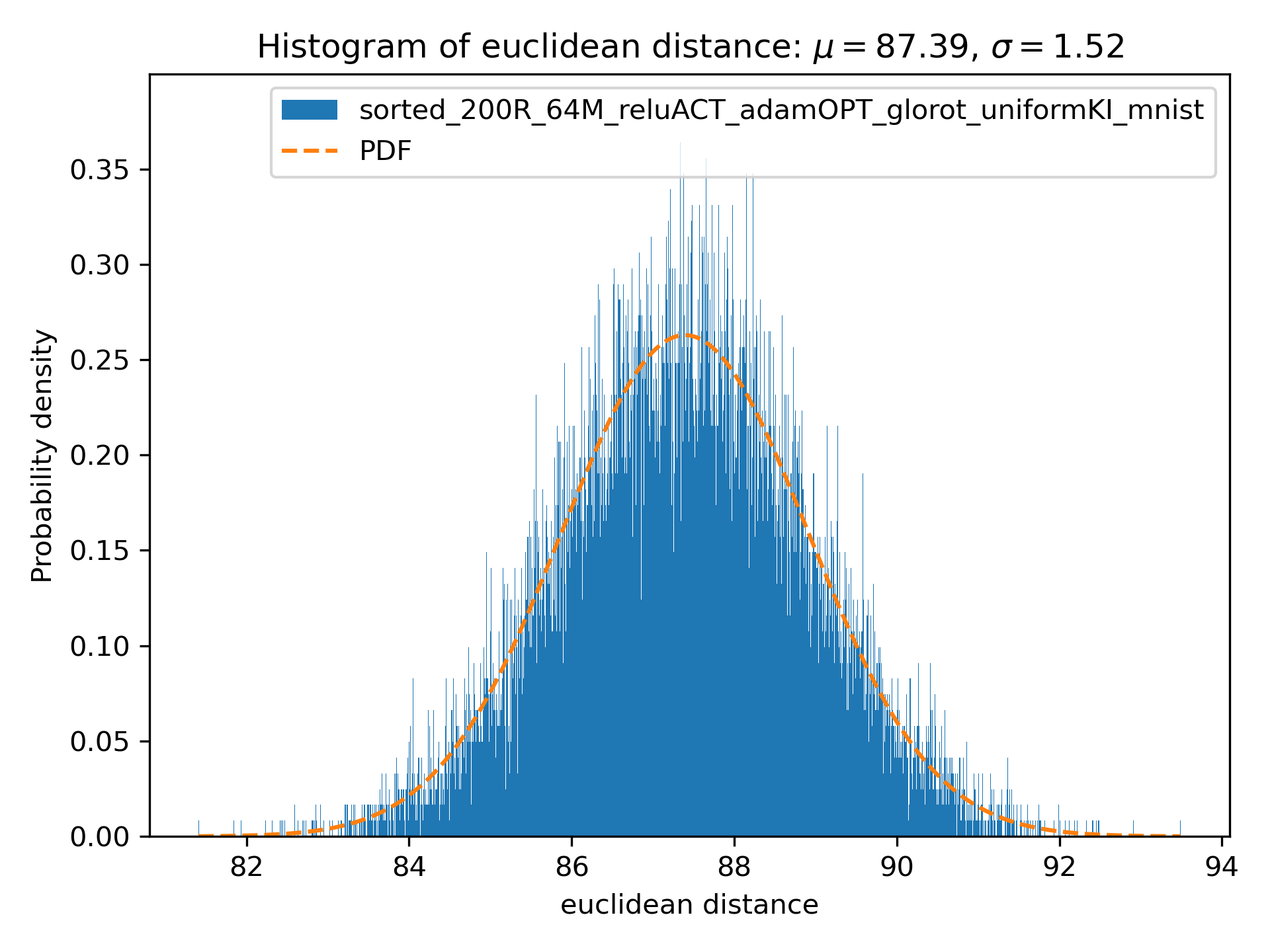}}
        \subfloat[Adam-256, $\mu=40.38$, $\sigma=0.17$]{\includegraphics[width=0.3\linewidth]{fig/adam3.png}}
        \quad
	\subfloat[Adam-512, $\mu=45.86$, $\sigma=0.11$]{\includegraphics[width=0.3\linewidth]{fig/adam4.png}}
        \subfloat[Adam-1024, $\mu=51.44$, $\sigma=0.07$]{\includegraphics[width=0.3\linewidth]{fig/adam5.png}}
        \subfloat[Adam Normalized distributions]{\includegraphics[width=0.3\linewidth]{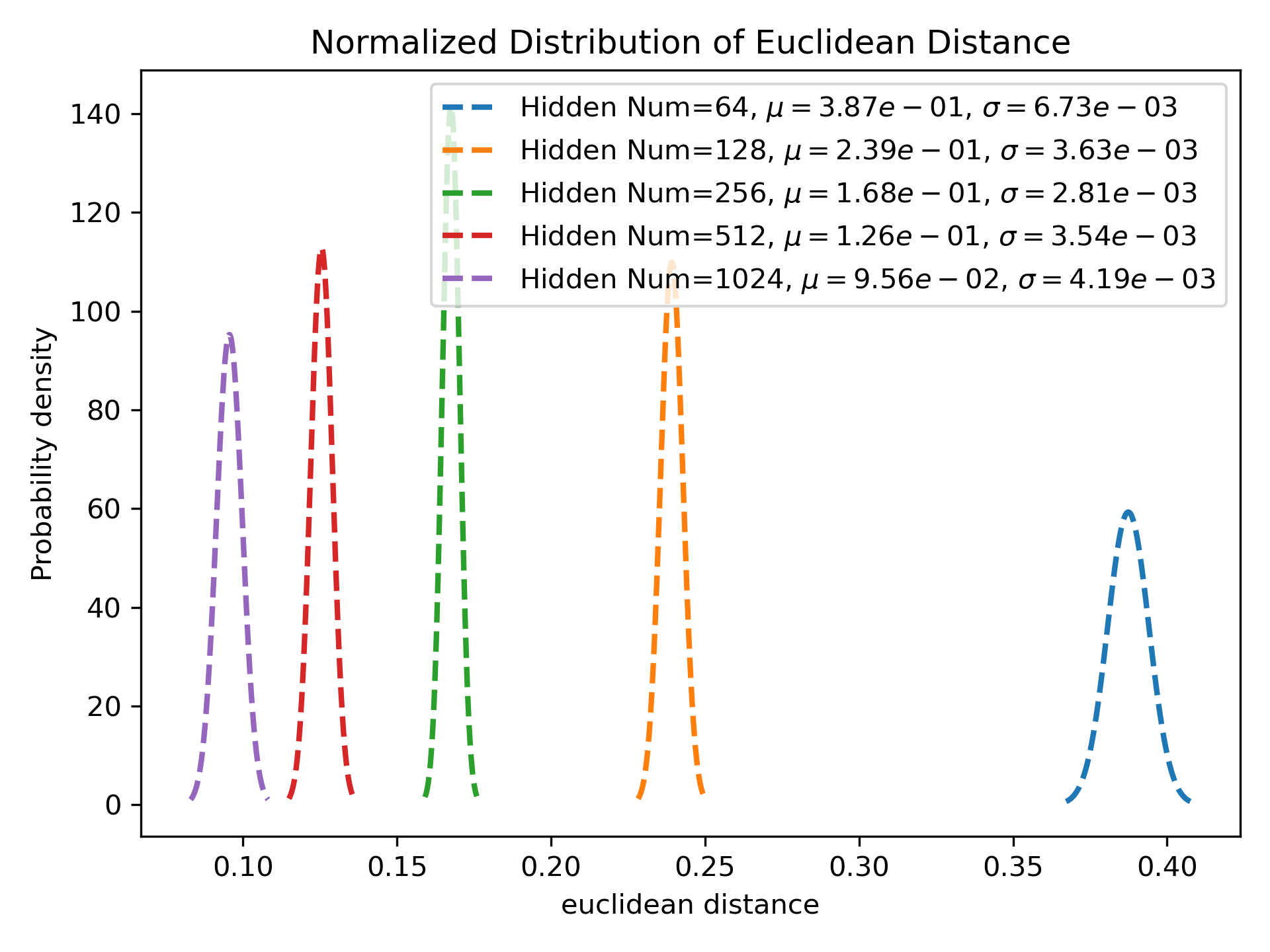}}
\caption{Experimental results for training a single hidden layer ReLU model with the Adam optimization algorithm on MNIST. Repeat $R=100$ for each parameter configuration. The horizontal axis represents the euclidean distance between each pair of solutions. The vertical axis denotes probability density. The initialization configuration is drawn from glorot-uniform. The number of nodes in the hidden layer are 64, 128, 256, 512 and 1024.}
\label{fig:exp_res3}
\end{figure*}

We trained a single hidden layer ReLU model on the MNIST dataset using the SGD and Adam optimization algorithm, with weights initially randomized from a glorot uniform distribution, repeated for 100 times.
These solutions were subsequently reordered using Hierarchy Clustering Algorithm to mitigate the effect of permutation symmetry in the hidden layers. 
The Euclidean distance between each pair of the trained solutions are computed.
A Gaussian
distribution was best-fitted to the distribution for comparison. 
Figure \ref{fig:exp_res3} shows the experimental results of a single hidden layer ReLU model with random weight initialization from glorot uniform distribution using Adam optimization algorithm on the MNIST dataset when the training loss is less than $1e-5$.
Each set of hyperparameters is repeated 200 times to obtain a set of 200 solutions with different weight configurations.
Then, each solution is sorted, and the Euclidean distance between each pair of solutions is calculated after eliminating the interference of arrangement symmetry.
Then, the result is divided into $1,990$ intervals according to the upper and lower limits $[euDist_{min},euDist_{max}]$, the number of points in each interval is counted, a histogram is drawn based on this data, and the obtained histogram is fitted with Gaussian distribution.
The figure shows the results for the $FC_1-64$ through $FC_1-1024$ networks, organized from left to right and top
to bottom.

\begin{figure*}[t]
	\small
	\centering
        \subfloat[SGD-64, $\mu=32.64$, $\sigma=0.42$]{\includegraphics[width=0.3\linewidth]{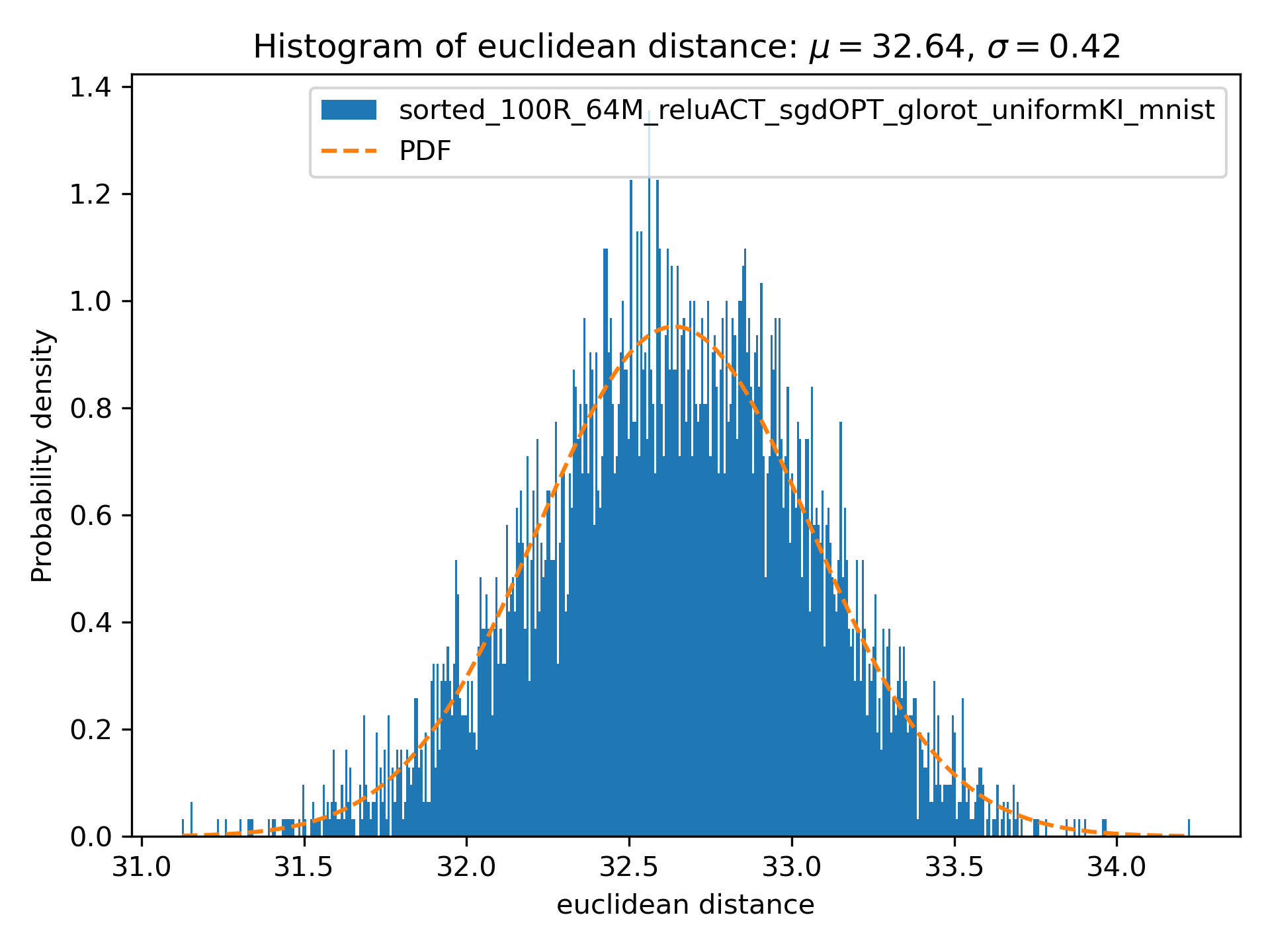}}
        \subfloat[SGD-128, $\mu=35.90$, $\sigma=0.27$]{\includegraphics[width=0.3\linewidth]{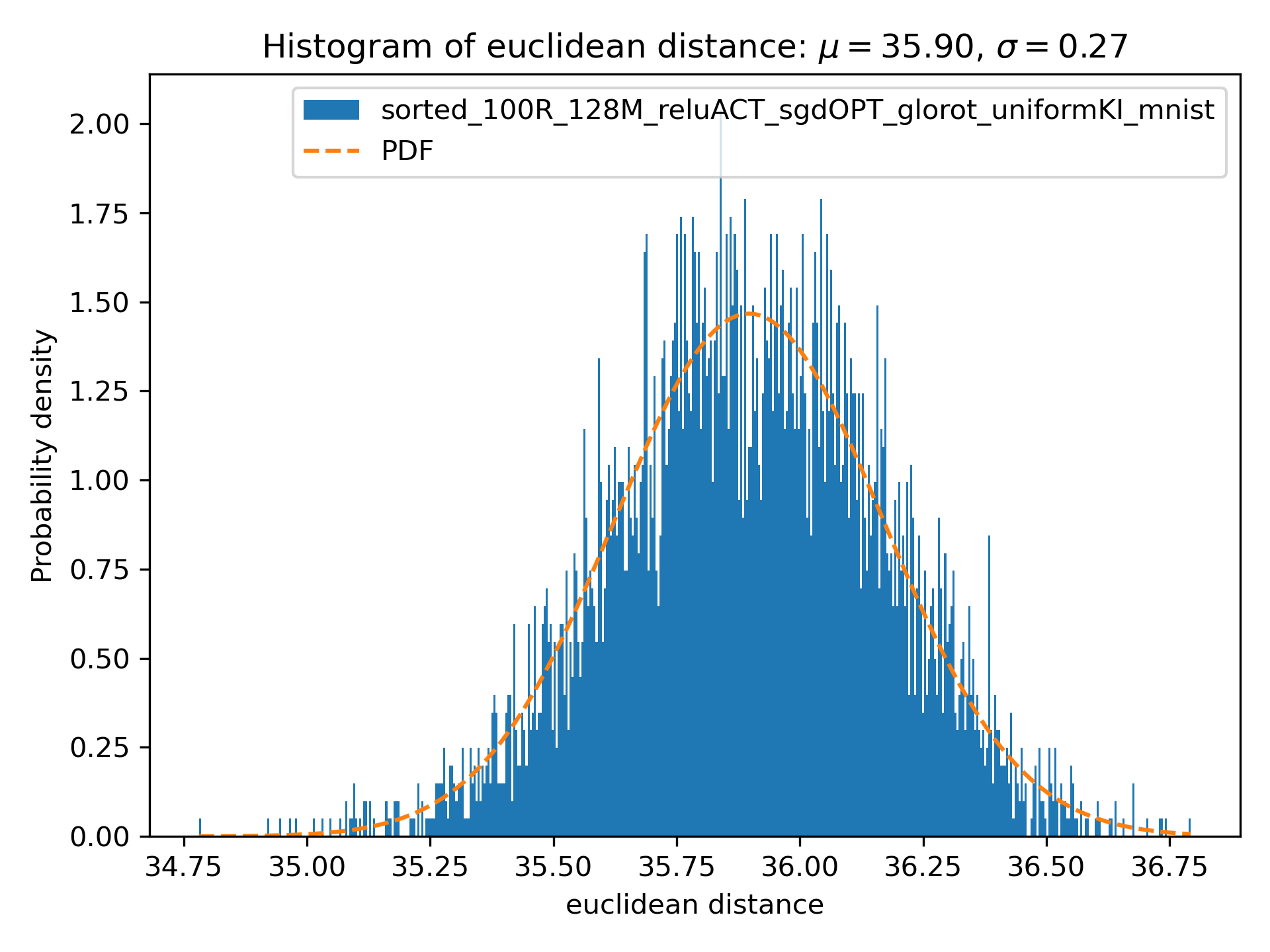}}
        \subfloat[SGD-256, $\mu=40.38$, $\sigma=0.17$]{\includegraphics[width=0.3\linewidth]{fig/sgd3.png}}
        \quad
	\subfloat[SGD-512, $\mu=45.86$, $\sigma=0.11$]{\includegraphics[width=0.3\linewidth]{fig/sgd4.png}}
        \subfloat[SGD-1024, $\mu=51.44$, $\sigma=0.07$]{\includegraphics[width=0.3\linewidth]{fig/sgd5.png}}
        \subfloat[SGD Normalized distributions]{\includegraphics[width=0.3\linewidth]{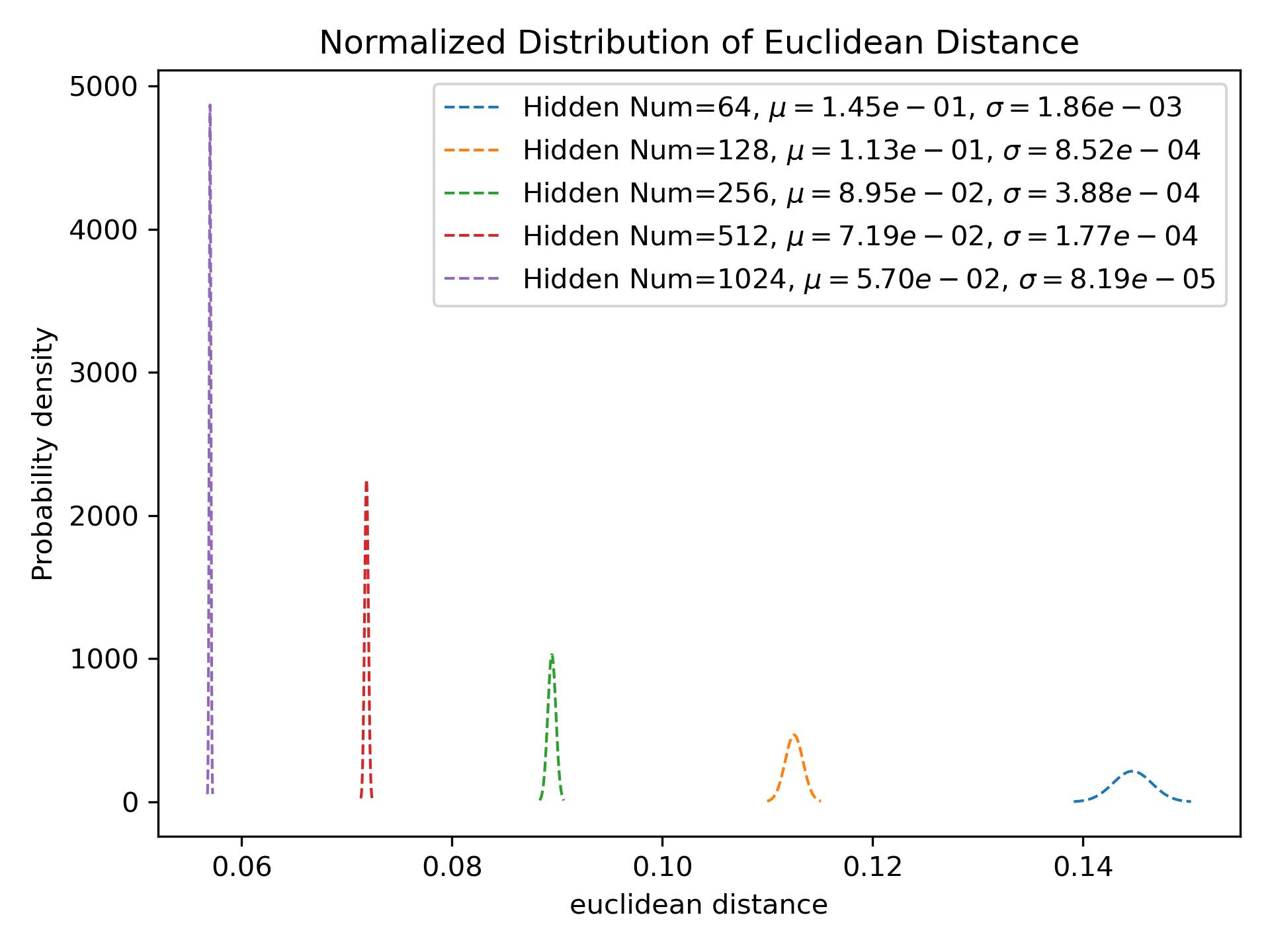}}
\caption{Experimental results for training a single hidden layer ReLU model with the Adam optimization algorithm on MNIST. The other settings are the same.}
\label{fig:exp_res6}
\end{figure*}

From Figure \ref{fig:exp_res6}, we can know that due to the characteristics of the SGD optimization algorithm, the Euclidean distance of the final solution trained from different initial points in the weight parameter space roughly conforms to the Gaussian distribution. In other words, the Euclidean distance, or similarity index, of most solutions is concentrated in $[\mu-\sigma, \mu+\sigma]$. And as the number of hidden layer nodes increases, the mean value $\mu$ gradually increases and the standard deviation $\sigma$ gradually decreases. Reflected on the loss landscape, it means that in this training scenario, the solutions are concentrated in a small area. And as the number of parameters increases, the phenomenon of concentration becomes more and more obvious. A good judgment cannot be made on the question of whether there are multiple extreme values in the loss landscape.

Compared with Figure \ref{fig:exp_res3}, we can see that, due to the characteristics of the Adam optimization algorithm, when the number of hidden layer nodes is less than 256, the results are basically consistent with the SGD algorithm.
When the number of hidden layer nodes is 512 and 1024, the Euclidean distance between different solutions no longer conforms to the Gaussian distribution, and a multi-peak phenomenon appears.
Especially when the number of nodes is 1024, there are five clear peaks.
Reflected in the loss landscape, it means that in this training scenario, different solutions are concentrated in multiple regions, and as the number of parameters increases, the concentration phenomenon becomes more and more obvious.
At this point, we can make a preliminary inference that there are multiple different local minima in the loss landscape of the single hidden layer ReLU network, and when the number of parameters in the network meets a certain condition, the Adam optimization algorithm can be used to train to obtain solutions located in different local minima.

\begin{figure*}[tb]
	\small
	\centering
        \subfloat[$M=64$]{\includegraphics[width=0.2\linewidth]{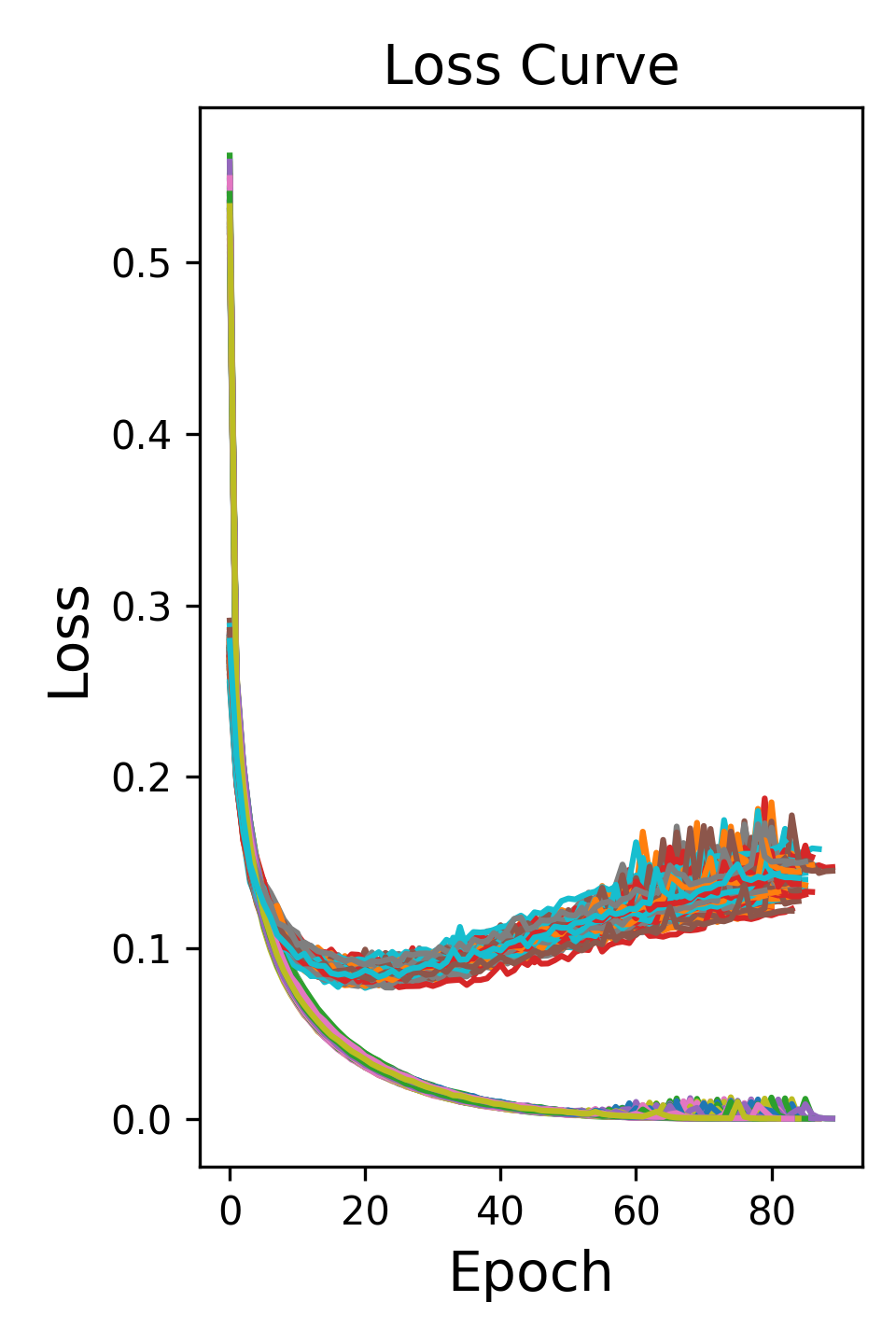}}
        \subfloat[$M=128$]{\includegraphics[width=0.2\linewidth]{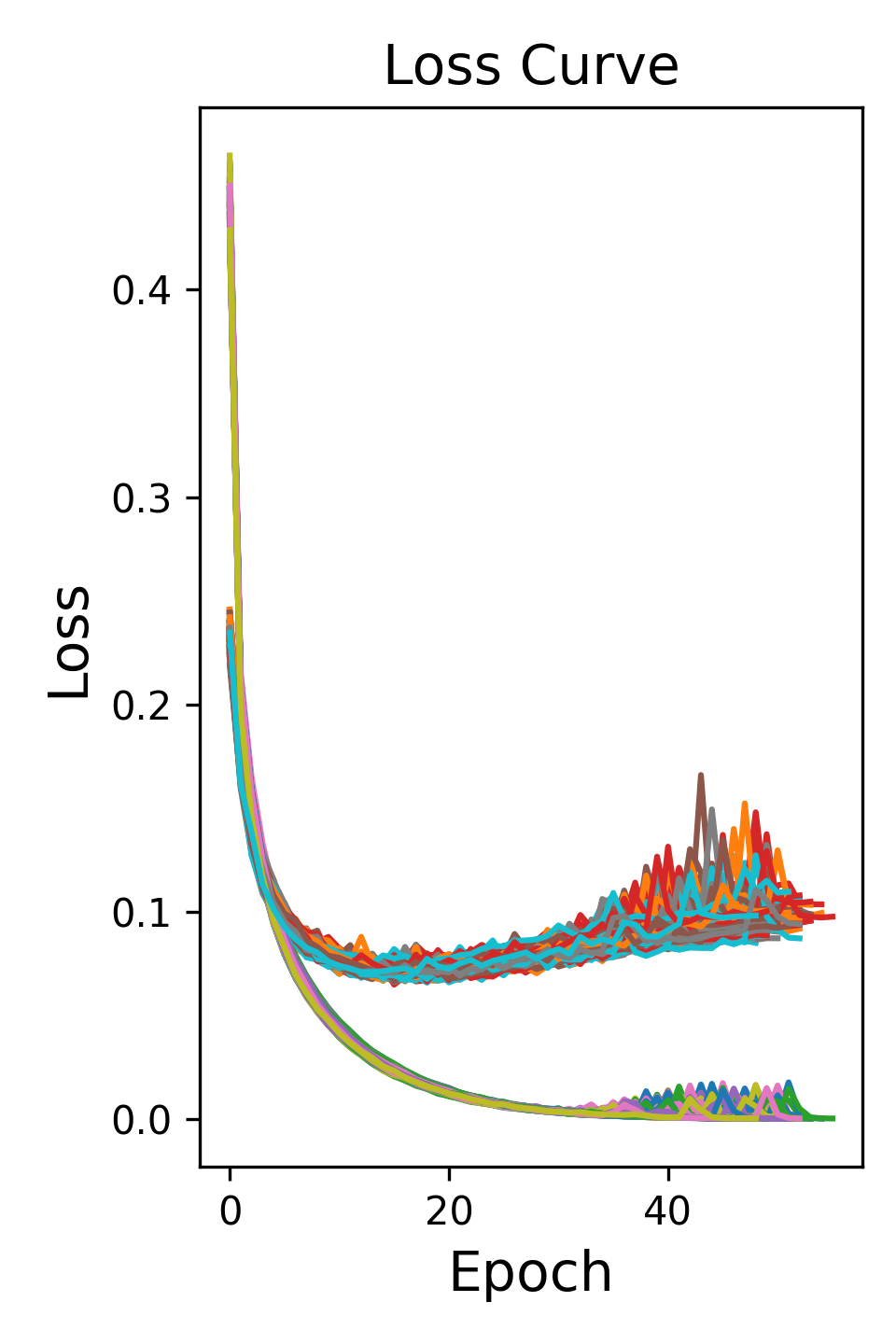}}
        \subfloat[$M=256$]{\includegraphics[width=0.2\linewidth]{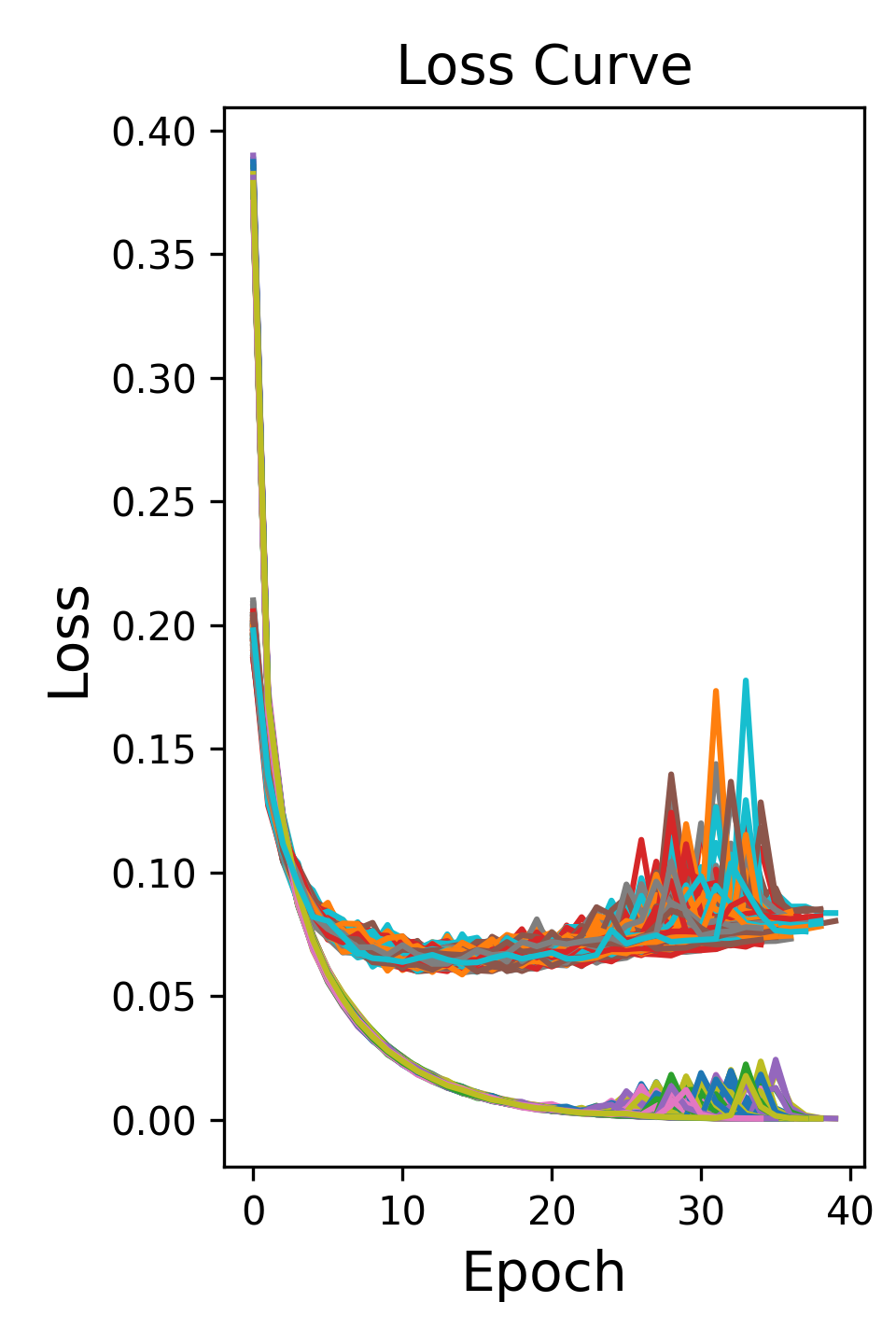}}
        \subfloat[$M=512$]{\includegraphics[width=0.2\linewidth]{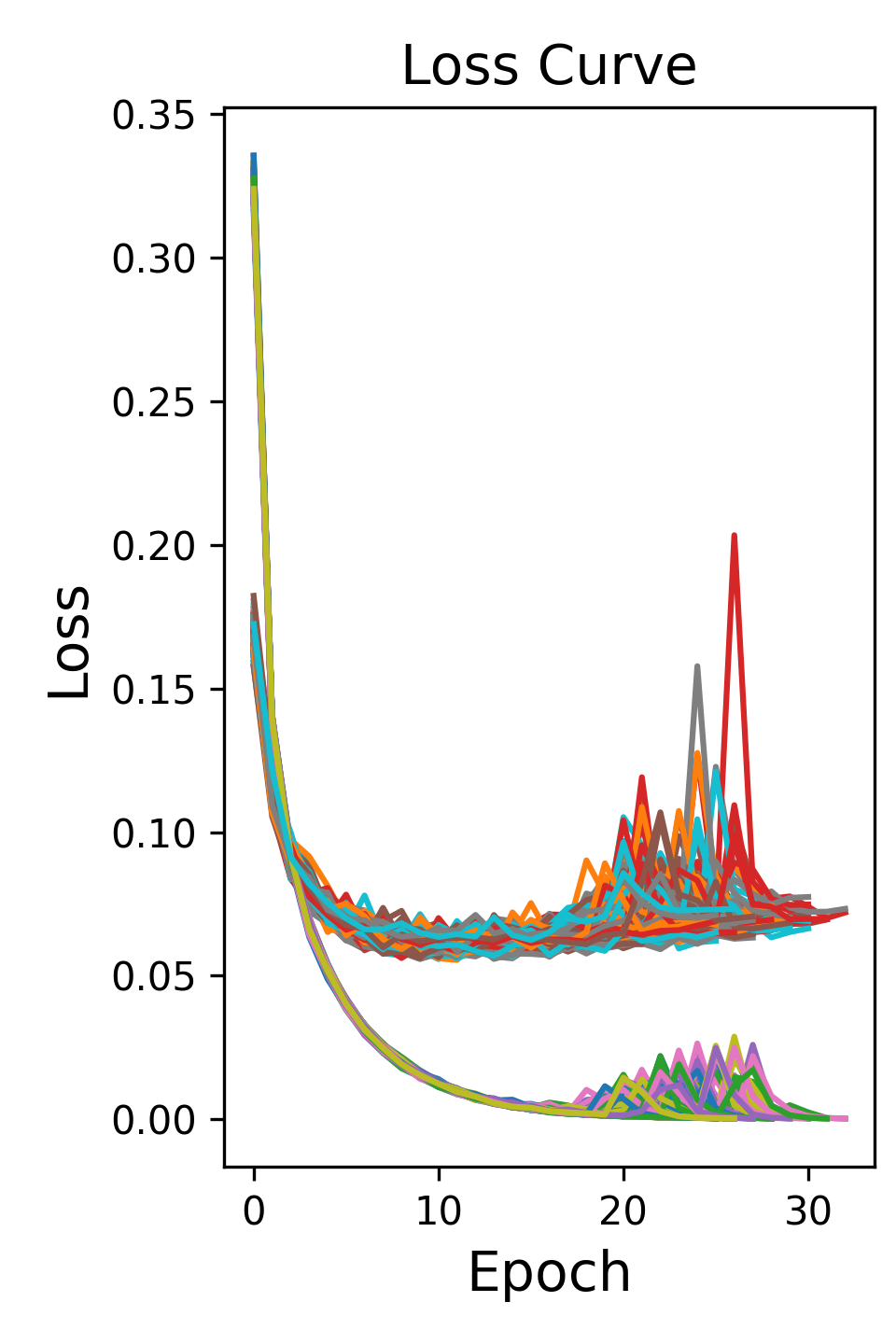}}
        \subfloat[$M=1024$]{\includegraphics[width=0.2\linewidth]{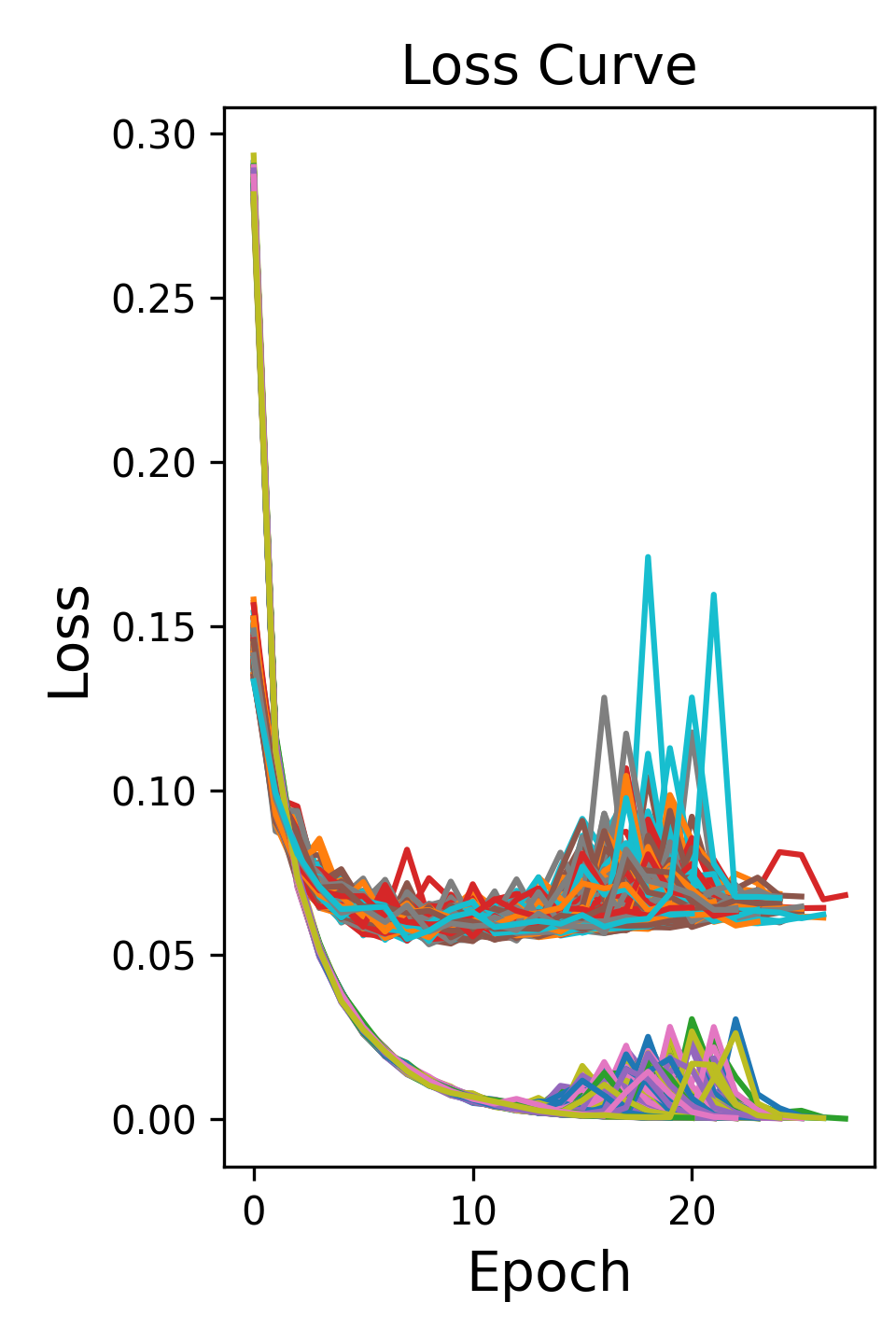}}
        \quad
	\subfloat[$M=64$]{\includegraphics[width=0.2\linewidth]{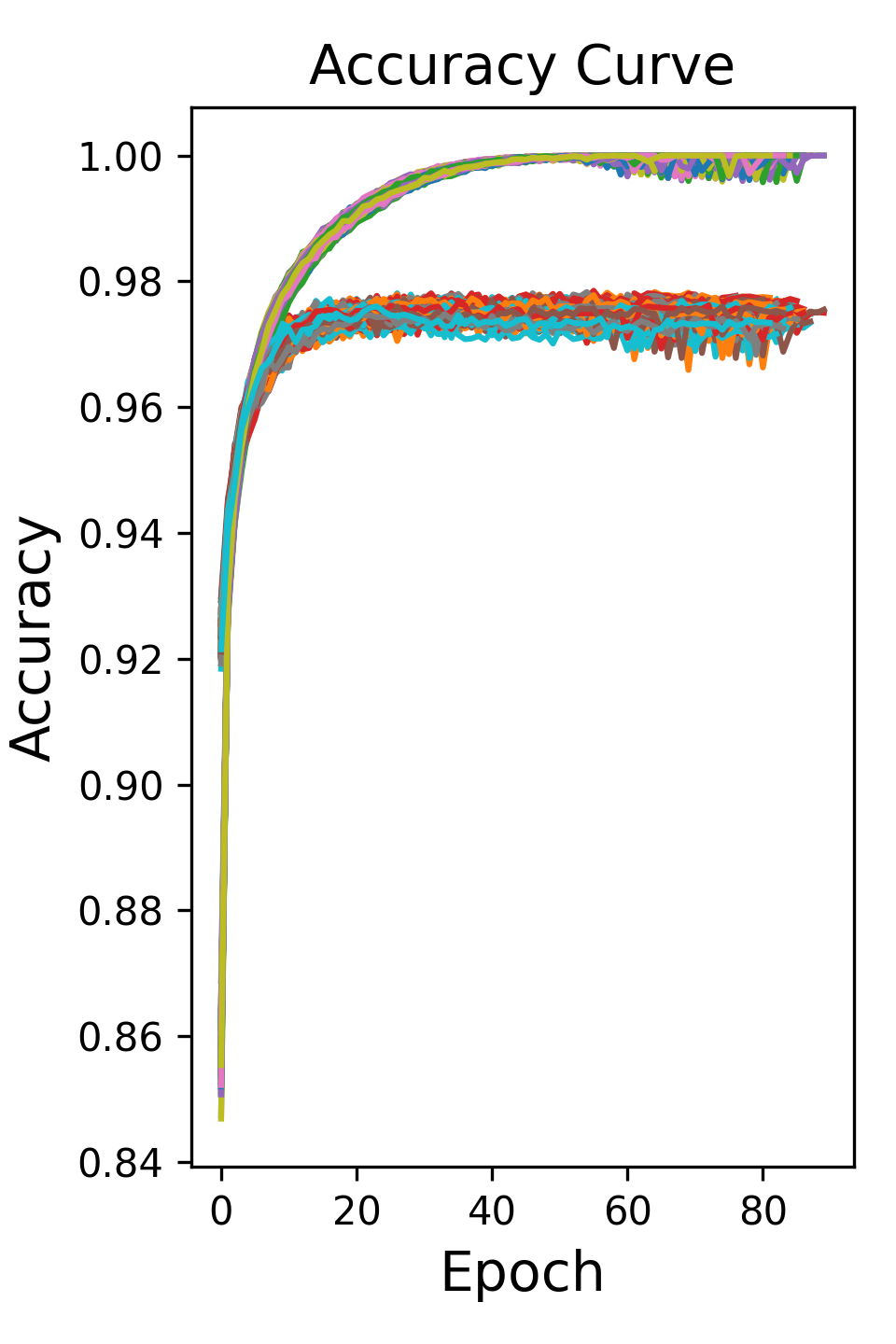}}
        \subfloat[$M=128$]{\includegraphics[width=0.2\linewidth]{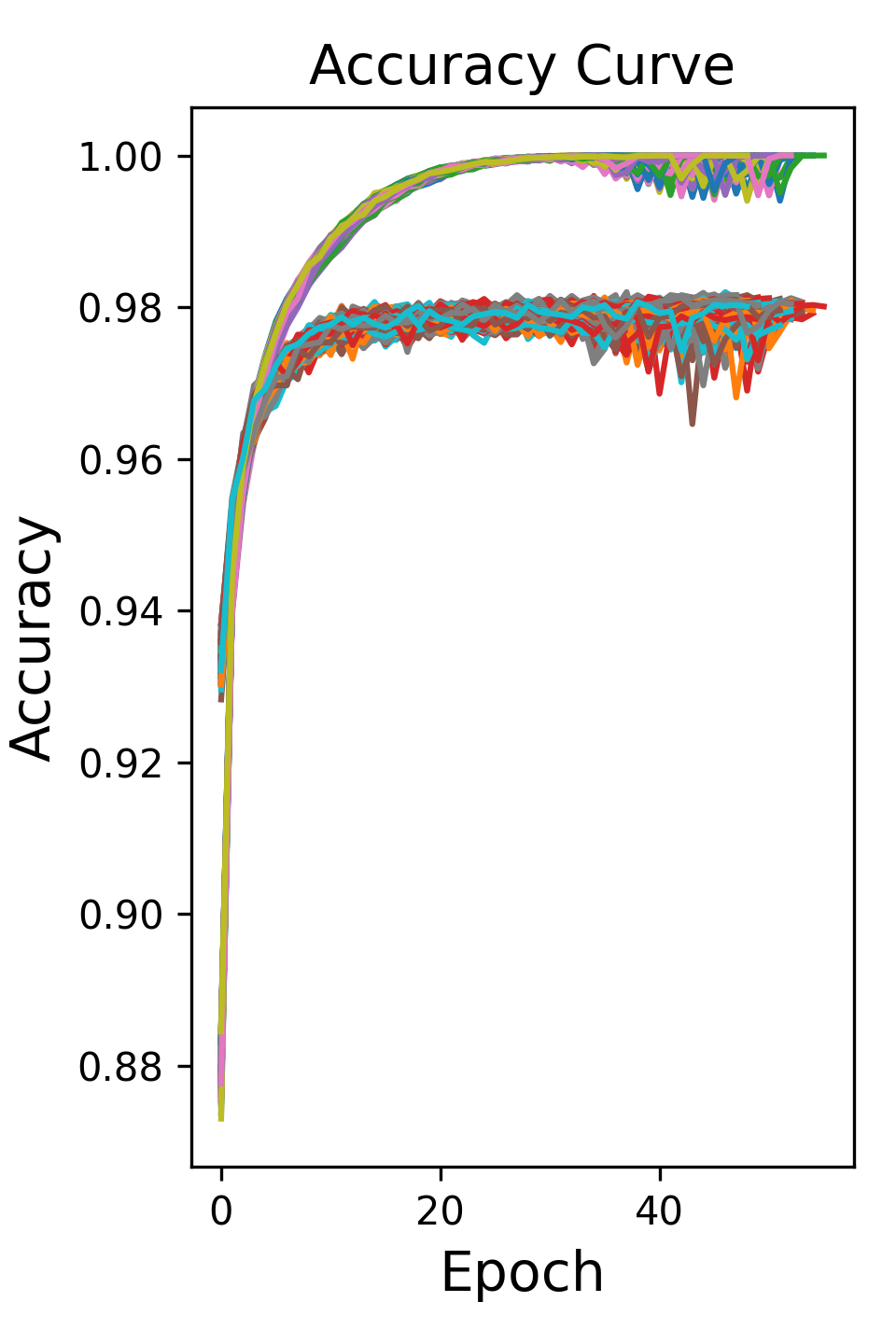}}
        \subfloat[$M=256$]{\includegraphics[width=0.2\linewidth]{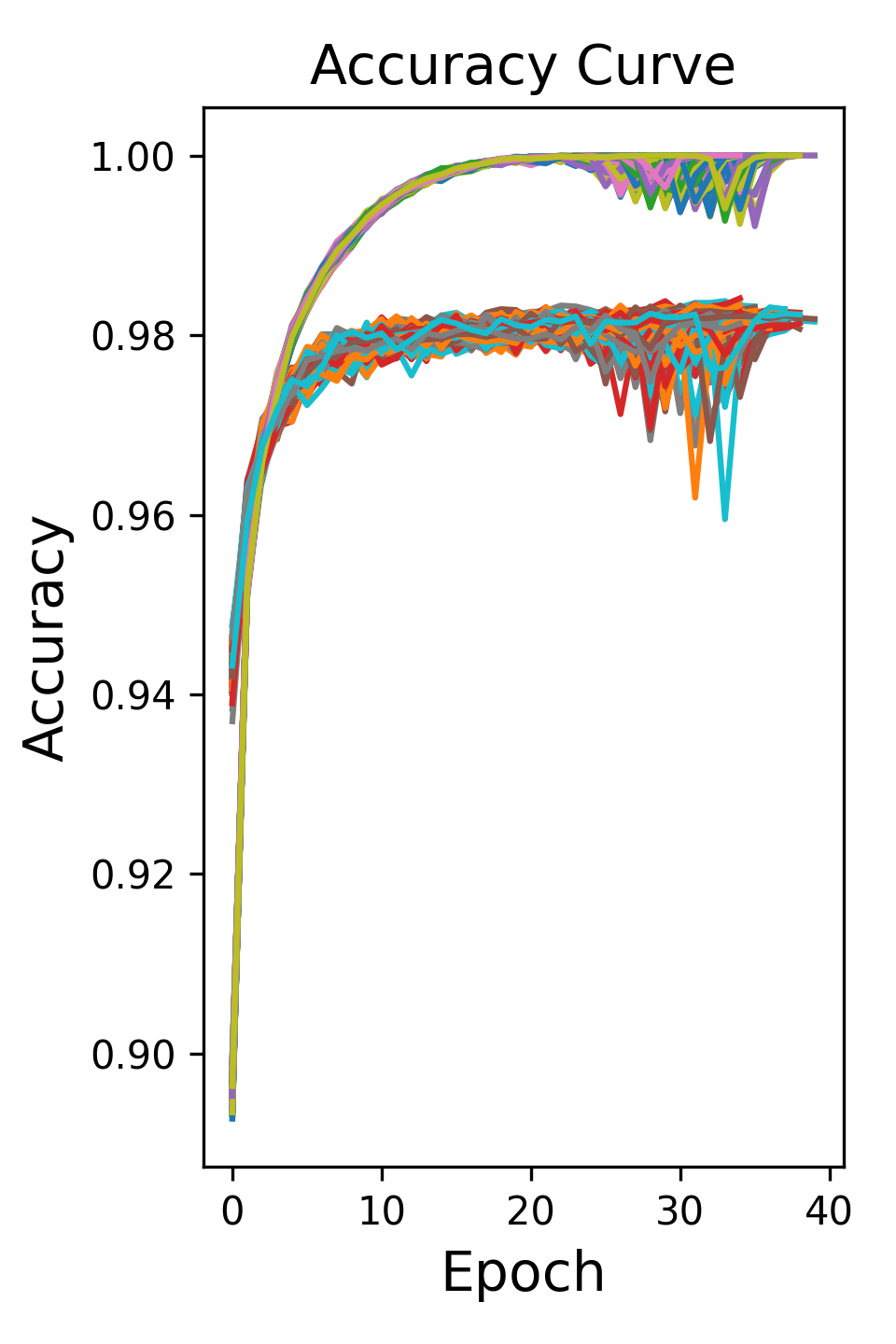}}
        \subfloat[$M=512$]{\includegraphics[width=0.2\linewidth]{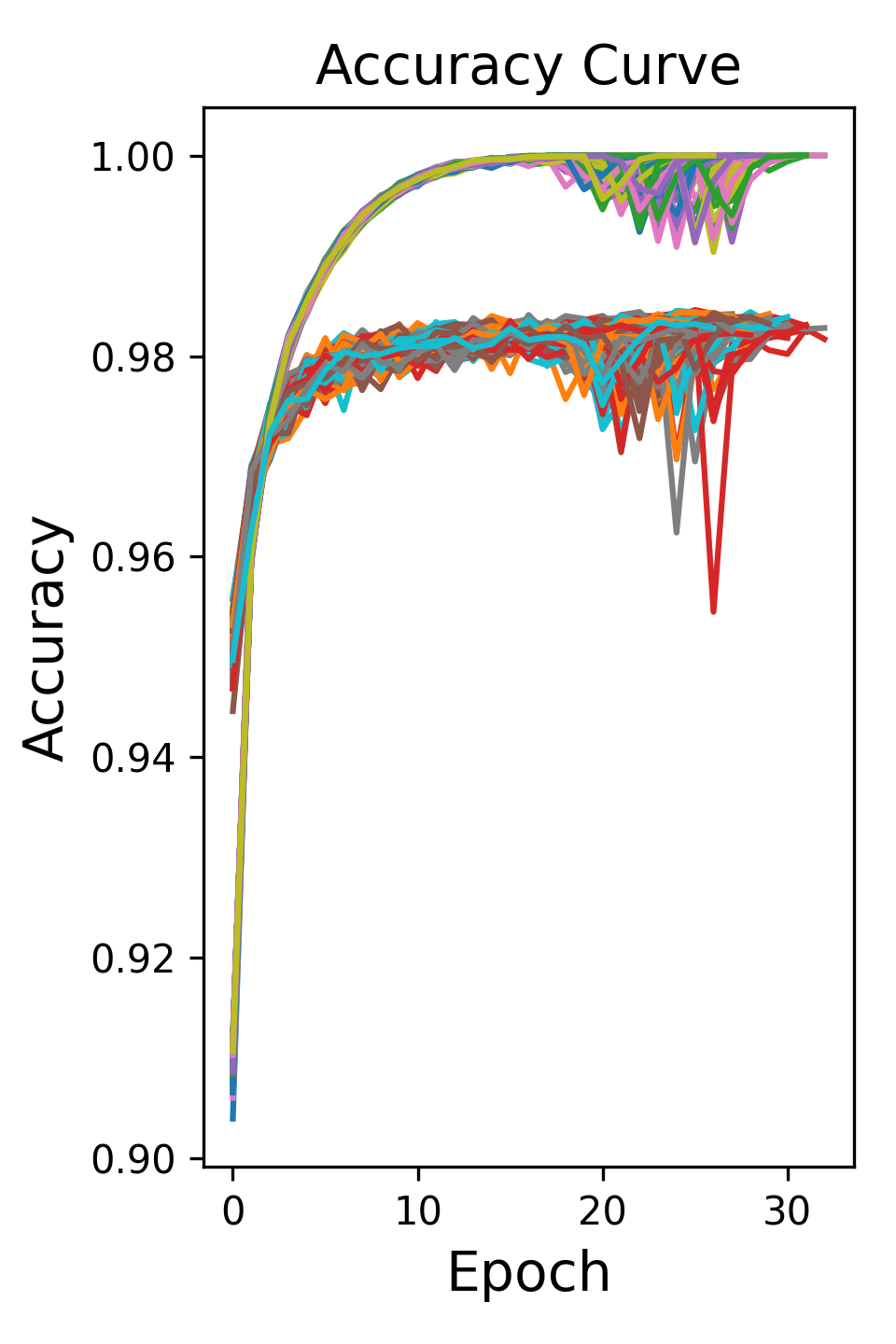}}
        \subfloat[$M=1024$]{\includegraphics[width=0.2\linewidth]{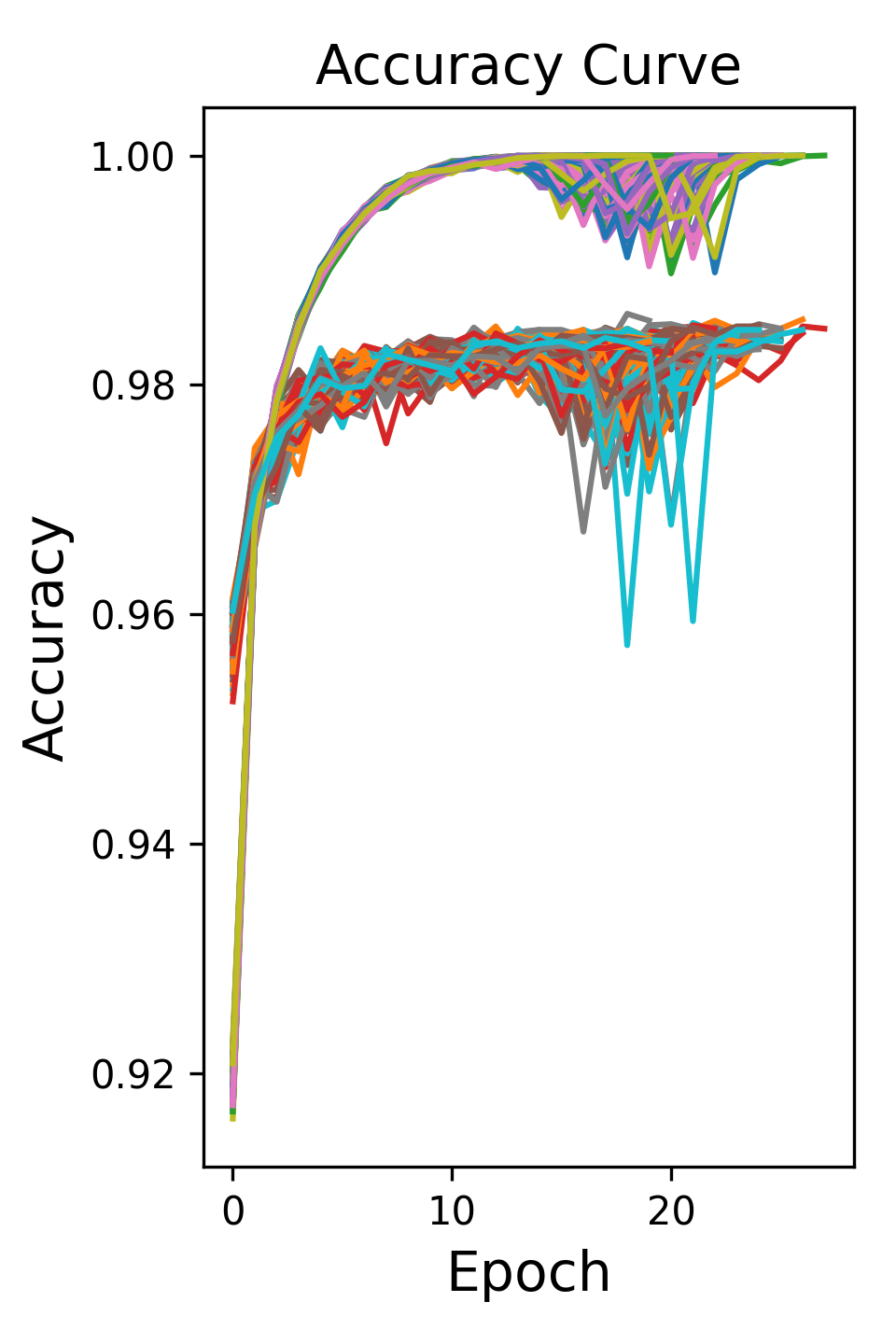}}
\caption{Experimental results of loss and accuracy on different neural networks. For each parameters, we repeated the experiments for $200$ times. In the Acc image of each parameter, the higher curve is the training accuracy and the lower curve is the test accuracy. The value of $M$ represents the nodes of middle layer.}
\label{fig:exp_res7}
\end{figure*}

We investigate the generalization of the model. Based on the loss and accuracy values recorded during the training process, images of loss and accuracy changing with the number of iterations can be made, as shown in Figure \ref{fig:exp_res7}.
The generalization performance of the model, that is, the rate of  accuracy on the test set, is an indicator for evaluating the performance of different local minima. 
It can be seen that in about 5 iterations, all networks can find solutions that are not much different from their own optimal solutions.
Later, as the number of iterations increases, the network model with a larger number of hidden layer nodes will have greater fluctuations, but at the end of the iteration, it will all reach optimal performance.

\subsubsection{Comparison between Different Batch Size}

\begin{table}[tb]
    \centering
    \caption{Performance of SB and LB in four networks}
    \begin{tabular}{c|c|c|c|c}
    \toprule
    \multirow{2}{*}{\textbf{\begin{tabular}[c]{@{}l@{}} Network \\ Structure\end{tabular}} } & \multicolumn{2}{c|}{\textbf{\begin{tabular}[c]{@{}l@{}} Training \\ Accuracy\end{tabular} } } & \multicolumn{2}{c}{\textbf{\begin{tabular}[c]{@{}l@{}} Testing \\ Accuracy\end{tabular} } } \\ \cline{2-5}
    & \textbf{SB} & \textbf{LB} & \textbf{SB} & \textbf{LB} \\ \hline
    $FC_1$-512 & $99.99$ & $99.99$ & $98.31$ & $97.56$ \\
    $FC_1$-1024 & $99.99$ & $99.99$ & $98.69$ & $97.97$ \\
    $FC_1$-2048 & $99.99$ & $99.99$ & $98.78$ & $98.10$ \\
    $FC_1$-4096 & $99.99$ & $99.99$ & $98.99$ & $98.49$ \\
    \bottomrule
    \end{tabular}
    \label{tab:SB_LB}
\end{table}

The experimental results represented by $FC_1$-512 are in Table \ref{tab:SB_LB}. It can be found that the training accuracy has reached $99.99\%$, indicating that these four networks can basically fully represent the MNIST data set. 
Moreover, the test accuracy of SB (Small Batch) in the four networks is higher than that of LB (Large Batch), and the average accuracy has improved by $0.67\%$.

Despite this, the generalization performance of the four networks did not exceed $99.00\%$. Through observation, we can find that the generalization performance can be improved to a certain extent by increasing the number of hidden layer nodes.

Due to space limitations and the similarity of experimental results, only the experimental results on the MNIST data set are shown here, see Table \ref{tab:SB_LB}.
Observing Table \ref{tab:SB_LB}, we can find that the training accuracy reaches 99.99\%, indicating that these four networks can basically fully represent the MNIST data set. However, the generalization performance of the four networks does not reach above 99.00\%.
And increasing the number of hidden layer nodes can improve the generalization performance to a certain extent. The test accuracy of SB in the four networks is higher than that of LB, with an average accuracy improvement of 0.67\%.

\begin{figure*}[t]
    \centering
    \includegraphics[width=\columnwidth]{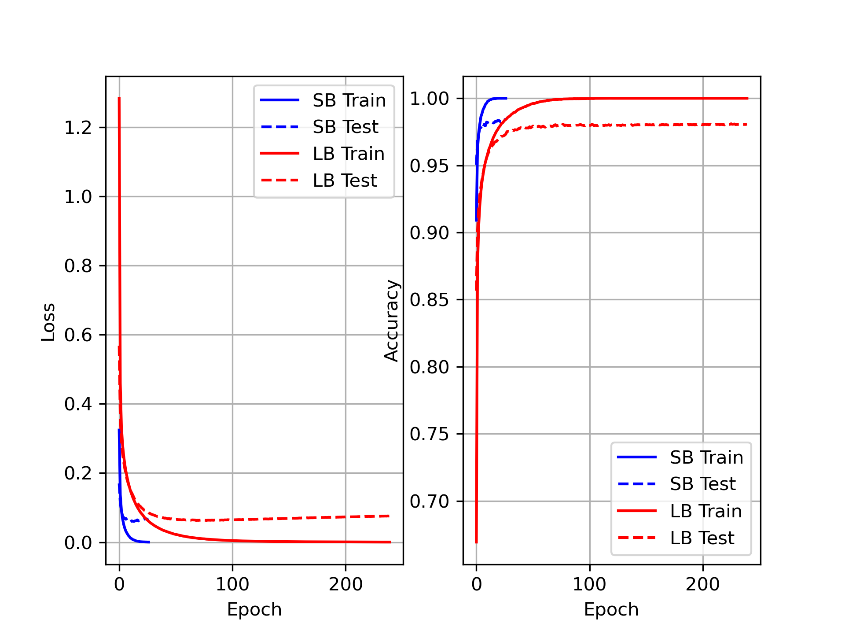}
    \caption{The training process curve of $FC_1$-512. SB denotes small mini-batch. LB denotes large mini-batch.}
    \label{fig:exp_res9}
\end{figure*}

The training and testing curve represented by $FC_1-512$ is shown in Figure \ref{fig:exp_res9}.

\end{document}